\documentclass[preprint,12pt]{aastex}

\newcommand{\degree}{\mbox{$^{\circ}$}}
\newcommand{\am}{\mbox{\arcmin}}
\newcommand{\as}{\mbox{\arcsec}}

\newcommand{\msun}{\mbox{M$_\odot$}}

\input{epsf}

\def\plotfiddle#1#2#3#4#5#6#7{\centering \leavevmode
\vbox to#2{\rule{0pt}{#2}}
\includegraphics{#1}}

\begin{document}

\slugcomment{v3.0; 28 May 2012}

\renewcommand\thefootnote
	{\fnsymbol{footnote}}
\title {\bf A  {\it Herschel} Survey of Cold Dust in Disks Around Brown Dwarfs and Low-Mass Stars\footnote{{\it Herschel} is an ESA space observatory with science instruments provided by European-led Principal Investigator consortia and with important participation from NASA.}}
\author{Paul M. Harvey\altaffilmark{1},
Thomas Henning\altaffilmark{2},
Yao Liu\altaffilmark{3,4},
Fran\c cois M\'enard\altaffilmark{5},
Christophe Pinte\altaffilmark{5},
Sebastian Wolf\altaffilmark{3},
Lucas A. Cieza\altaffilmark{6},
Neal J. Evans II\altaffilmark{1},
Ilaria Pascucci\altaffilmark{7}
}

\altaffiltext{1}{Astronomy Department, University of Texas at Austin, 1 University Station C1400, Austin, TX 78712-0259;  pmh@astro.as.utexas.edu, nje@astro.as.utexas.edu}
\altaffiltext{2}{Max Planck Institut for Astronomy, K\"onigstuhl 17, 69117 Heidelberg, Germany;henning@mpia.de}
\altaffiltext{3}{University of Kiel, Institute of Theoretical Physics and Astrophysics, Leibnizstr. 15, 24098 Kiel, Germany; wolf@astrophysik.uni-kiel.de, yliu@pmo.ac.cn}
\altaffiltext{4}{Graduate School of the Chinese Academy of Sciences, Beijing 100080, China; yliu@pmo.ac.cn}
\altaffiltext{5}{UJF-Grenoble 1/CNRS-INSU, Institut de Plan\'etologie et d'Astrophysique (IPAG) UMR 5274, BP 53, 38041 Grenoble cedex 9, France; menard@obs.ujf-grenoble.fr, christophe.pinte@obs.ujf-grenoble.fr}
\altaffiltext{6}{Institute for Astronomy, University of Hawaii, 2680 Woodlawn Dr., Honolulu, HI 96822; lcieza@ifa.hawaii.edu, Sagan Fellow}
\altaffiltext{7}{Lunar and Planetary Laboratory, Department of Planetary Sciences, University of Arizona, 1629 E. University Blvd., Tucson, AZ 85721; pascucci@lpl.arizona.edu}

\begin{abstract}

We report the complete photometric results from our {\it Herschel} study which is the
first comprehensive program to search for far-infrared 
emission from cold dust around 
young brown dwarfs.  We surveyed 50 fields containing 51 known or suspected brown dwarfs and very low
mass stars that have evidence of circumstellar
disks based on {\it Spitzer} photometry and/or spectroscopy.  The objects with known spectral types range from M3 to M9.5.  
Four of the candidates were subsequently identified as 
extragalactic objects. 
Of the remaining 47 we have successfully detected 36 at 70\micron\ and 14 at 160\micron\ with S/N greater than 3, as well as several additional
possible detections with low S/N.  
The objects exhibit a range of [24]--[70] micron colors
suggesting a range in mass and/or structure of the outer disk.  We present modeling of the spectral energy distributions
of the sample and
discuss trends visible in the data.
Using two Monte Carlo radiative transfer codes we investigate
disk masses and geometry.
We find a very wide range in modeled total disk masses from less than $10^{-6}$ \msun\ up to $10^{-3}$ \msun\ with a median disk
mass of order $3 \times 10^{-5}$ \msun, suggesting that the median ratio of disk mass to central object mass may be lower
than for T Tauri stars.
The disk scale heights and flaring angles, however, cover a range consistent with those seen around T Tauri stars.
The host clouds in which the
young brown dwarfs and low-mass stars are located span a range in estimated age from $\sim$1--3 Myr to $\sim$10 Myr and represent a
variety of star-forming environments.
No obvious dependence on cloud location or age is seen in the disk properties, though the statistical significance of
this conclusion is not strong.

\end{abstract}


\section{Introduction}\label{intro}

The existence of circumstellar disks around low-mass stars and substellar objects is now a well-established
phenomenon.  Many of the observations of the lowest mass objects have taken advantage of the {\it Spitzer Space Telescope's} \citep{werner04}
exquisite sensitivity in the 3 -- 24\micron\ spectral region, 
e.g. \citet{allers06,hern07,guieu07,luhman05} (and many of the other references in Table \ref{aortable}) to find thermal emission above that
expected from the photospheres of these objects.  With the likely surface density gradient of such disks, however,
most of the dust is typically located at radii too large to emit strongly at those {\it Spitzer}
wavelengths, particularly for low-luminosity sub-stellar objects.  
At longer wavelengths, 70 -- 1000\micron, few disks have been
detected around brown dwarfs, e.g. \citep{klein03,scholz06,bouy08,riaz08}.  In addition to being more sensitive to the total disk mass, observations
at the longer wavelengths are also sensitive to the disk scale height, dust settling, and degree of flaring.  
The relationship between disk mass and structure and planet formation is highly uncertain, but
nominally one might expect objects with lower mass disks to form planetary systems with lower average mass planets \citep{mordasini12}.
Below some disk mass limit, though, there may be too little material for any planet formation, so understanding disks
around the lowest mass objects will place important constraints on likely planet formation around them.

The relatively
small number of detections of disks around very low-mass stars and brown dwarfs beyond the range of {\it Spitzer's}
24\micron\ MIPS band and its IRS instrument, has made it difficult to investigate the properties and
evolution of the cool dust component of these disks where most of the mass is located. 
We describe here the results from a {\it Herschel} GT1 (Guaranteed Time, Phase 1) program that obtained 70 and 160\micron\ photometry
of a statistically significant number of confirmed or candidate brown dwarfs and low-mass stars close to
the sub-stellar limit.  Our early results on the first three objects observed are described by \citet{harvey12} (hereafter, Paper I).
In the following sections we describe the observed sample, the details of the observing parameters and data reduction, the overall
properties of the spectral energy distributions (SED's), and extensive radiative transfer  modeling of the SED's to 
extract physical properties of the disks.  We conclude with a discussion of the overall results from this study
based on the complete sample together with existing data in the literature on cold dust in young brown dwarf and T Tauri disks.
The most important goals of this program were to use the {\it Herschel} far-ir photometry to place useful limits on
the outer disk properties around brown dwarfs, in particular, the mass and disk geometry.
We also aimed at providing as complete as possible SED coverage for future characterization at longer wavelengths with the Atacama
Large Millimeter Array (ALMA).

\section{The Sample}

Our sample of 50 targets was chosen from the literature to include young objects at or below the sub-stellar
limit in the nearby regions of star formation listed in Table \ref{cloudtable}.  Both confirmed and candidate BD's were included.
Table \ref{cloudtable} lists rough ages for each of the clouds based on the range of measurements in the
literature.  Although there are often more precise numbers quoted for median cloud ages in the literature, the age spreads
observed around the median have led us to consider only two age bins in our sample, the clouds with ages $\le$ 5 Myr and
those with ages of order 10 Myr.
Because the sub-stellar limit is only roughly defined as a
function of spectral type and age, we included objects generally with spectral types later than M5 with the exception of
several in the range M3 -- M4.8.
The most important additional selection criterion was evidence for
circumstellar dust emission, likely a disk, as shown by 8--24\micron\ excess emission over the photosphere.
Our goal was to reach flux limits at 70\micron\ low enough that we would be likely to detect disk emission
based on the shorter wavelength {\it Spitzer} data or to set useful upper limits.  Our sample was specifically
trimmed to not include objects that might be bright enough to be detected in {\it Herschel's} large scale
shallow surveys of star-forming regions \citep{andre10}.  
The selection process involved ``predicting'' the 70\micron\ flux density based on the 8--24\micron\ color
and 24\micron\ brightness using typical SED's for brighter young T Tauri stars as a guide.  As might be
expected, the final results showed our rough ``predictions'' to be accurate to only $\pm$ a factor of 3.
An additional selection criterion was an attempt to select objects that did not lie in regions of strong,
structured diffuse emission as based on {\it Spitzer} 24 and 70\micron\ images when available.  Even with this
criterion, one object, CFHTWIR96, was completely hidden in much brighter diffuse emission at both observed
wavelengths, and a number of sources
were invisible at 160\micron\ against the Galactic background/foreground.

Because the sample included candidate BD's, a few objects
included in the observed sample have turned out not to be stellar or sub-stellar objects with disks, including four found to be
extragalactic and one that may be at an earlier evolutionary state with a substantial envelope or extragalactic (ChaJ11080397-7730382).
The four extragalactic objects include 2MASSJ16080175-3912316 and IRACJ16084679-3902074 from the study by
\citet{allen07} and found to be extragalactic by \citet{com09} and \citet{com11}.  The source 2MASSJ1129140-7546256 was found
by \citet{luhman08b} to exhibit emission lines characteristic of a galaxy.  The object 2MASSJ16084747-3905087 from \citet{allen07}
has been found to consist of two nearby objects, one of which peaks in the near-ir and the other at longer wavelengths by Luhman (private
comm.).  This is most obvious in the {\it Spitzer} IRAC band 1 (3.6\micron) image where the object is clearly double,
consistent with it being a blend between a star and extragalactic object.

\begin{figure}
\plotfiddle{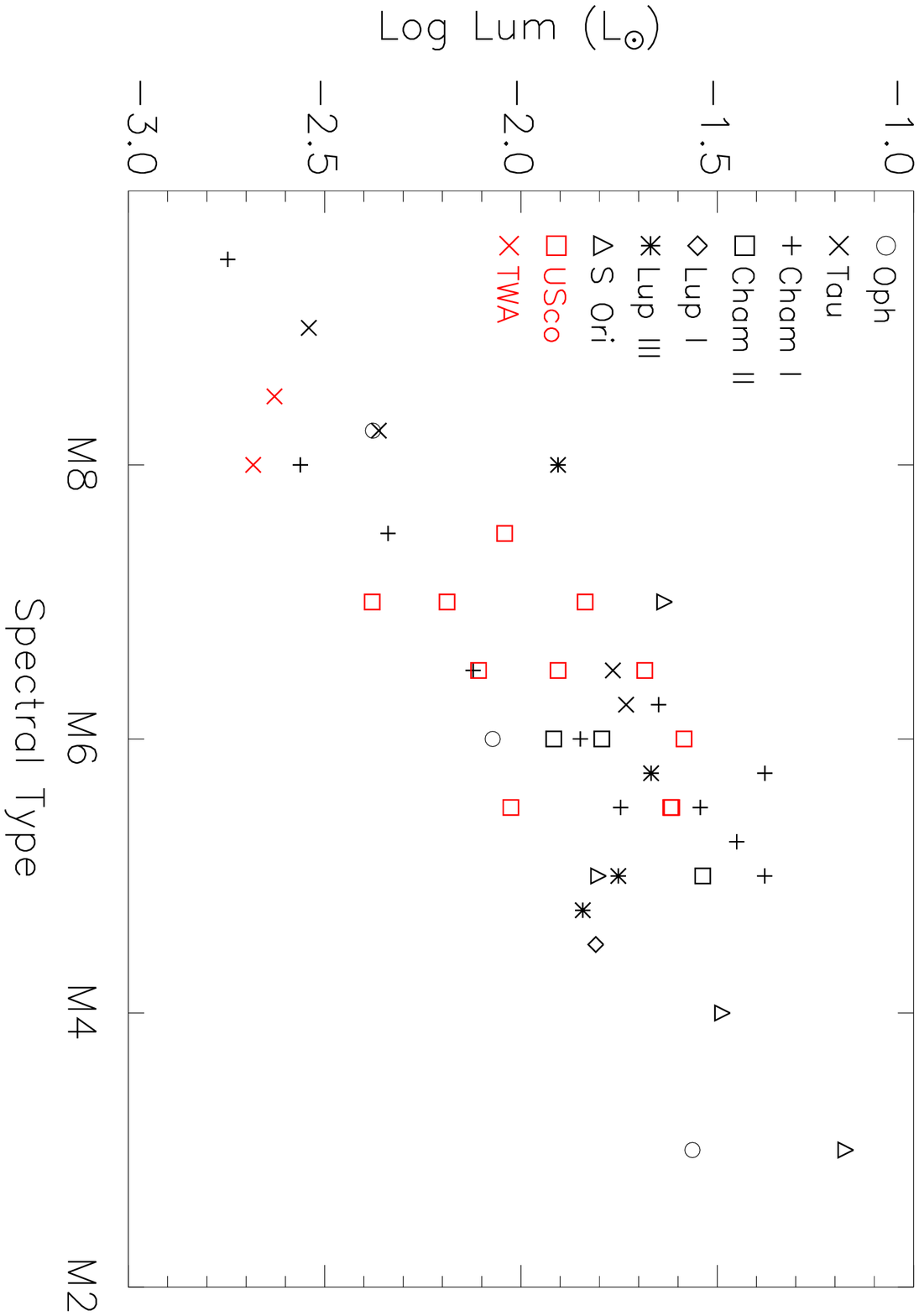}{6.5in}{90}{80}{80}{290}{70}
\figcaption{\label{lumspt}
Observed luminosities versus spectral types for the 43 objects with determined spectral types in the literature.  The
symbols are color coded so that the two star-forming regions that are significantly older than the others, U Sco and TWA, are in red.}
\end{figure}

\begin{figure}
\plotfiddle{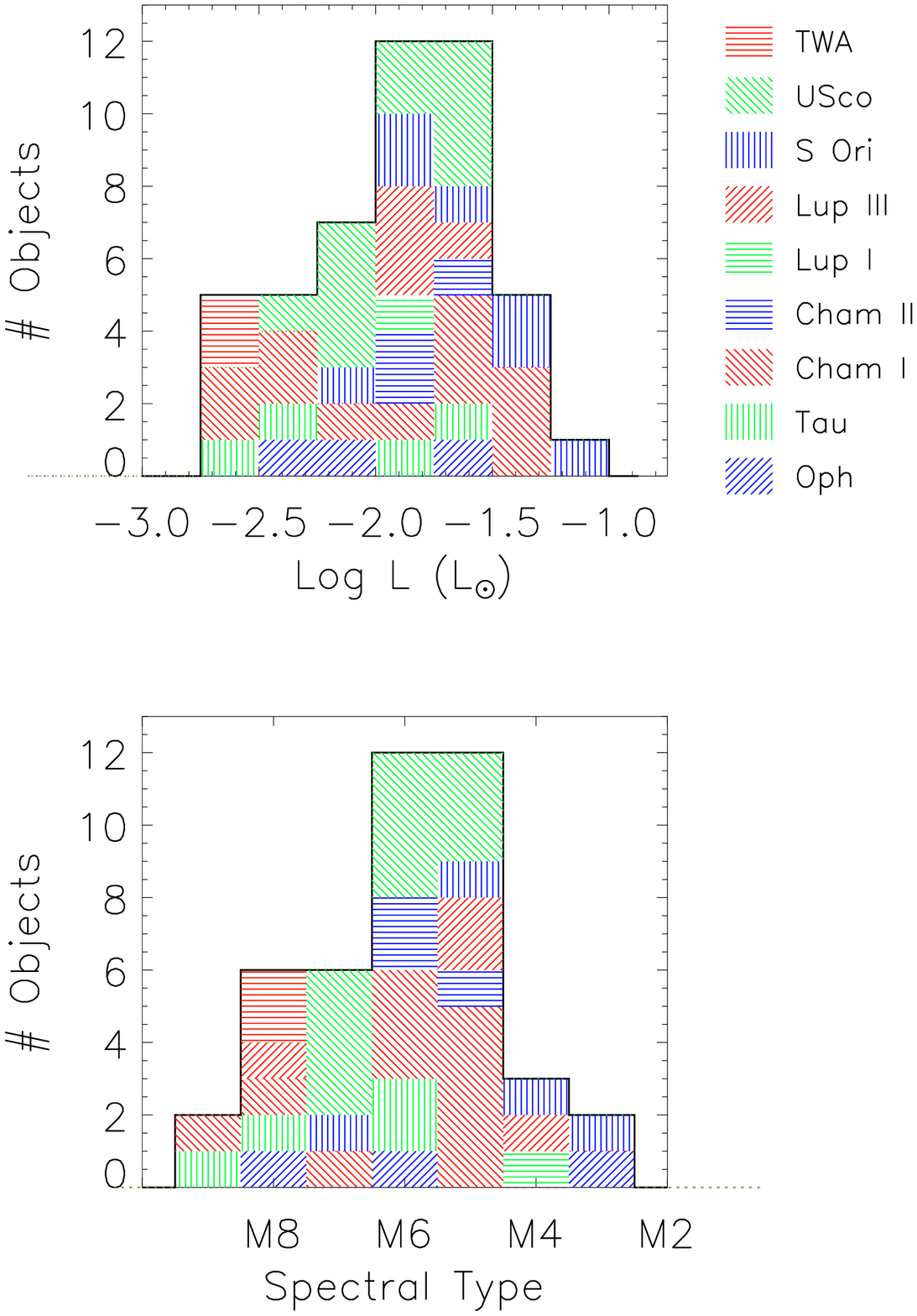}{6.5in}{0}{80}{80}{-180}{-30}
\figcaption{\label{lumhist}
Histogram of observed luminosities (1--200\micron) for all 47 objects and histogram of spectral types for the
43 objects with determined spectral types.}
\end{figure}

Table \ref{aortable} lists the complete sample of low-mass stars and brown dwarfs, including one object, ISO143 that was included serendipitously
in the field of ISO138 and which lies at the boundary between stellar
and sub-stellar objects.  With the inclusion of this object and the removal of the four extra-galactic sources,
the final sample consists of 46 likely or confirmed BD's/low mass stars with circumstellar disk emission, and one object,
ChaJ11080397-7730382 (hereafter cha110804), classified as a ``possible Class 1 source'' by \citet{luhman08a}.
We include this latter source because its {\it Spitzer} IRAC band 4 (8.0\micron) image shows faint nebulosity
around the object that is at least consistent with it being a Galactic source.  We have also measured upper limits to
its 1.6 and 2.2\micron\ fluxes from the 2MASS images of the area, since it was listed as undetected by the survey.
Figures \ref{lumspt} and \ref{lumhist} display the physical properties of the observed sample graphically.
Figure \ref{lumspt} shows a plot of the observed total luminosities (1--160\micron) versus spectral type for the 43 objects with
published spectral types assuming the distances listed in Table \ref{cloudtable}.  
This figure and several later figures show individual points with symbols for each host cloud; in addition, the two
oldest clouds, Upper Sco and TWA, are indicated with red symbols to highlight any possible age-dependent trends using
the two age bins discussed above.
Figure \ref{lumspt} illustrates graphically
the wide range in luminosities present in our sample.  It is also clear that there is a substantial range in
luminosity at any given spectral type and likewise a range in spectral types at any given luminosity.  There is no
obvious correlation with cloud age for either of these distributions.
Figure \ref{lumhist} shows histograms of the observed total luminosities of all the objects and the distribution
of spectral types for the 43 objects with published types; the separate contributions from each host cloud to
the total sample are indicated, illustrating the range of luminosity and spectral type present in the sample
from each host cloud.

Because of the various features of the above selection process, our study is not useful to estimate disk
frequencies for young brown dwarfs.  The sample, however, is useful to compare the frequencies of
cold dust emission relative to warmer dust emission due to, for example, differences in outer disk mass or
structure in the outer disk. More generally, our far-ir photometry has the potential to provide the first
statistically meaningful sample of disk masses around BD's.  Since we have objects from various star-forming regions covering a range in
estimated ages from $\sim$1--3 Myr  to $\sim$8--10 Myr, this sample may
also be useful for investigating disk evolution around BD's, e.g. mass and geometry.  We discuss all of these issues more fully in
\S \ref{disc}.

\section{Observations and Data Reduction}\label{obs}

Our program, GT1\_pharve01\_2, was designed to take advantage of the excellent sensitivity and spatial
sampling of the PACS photometer \citep{pacs10}  on {\it Herschel} \citep{pilbrat10}.  All the observations were made using exactly the same mode
with essentially the same total integration time.   We utilized the PACS ``mini-scan-map'' mode which images simultaneously at 
70 and 160\micron\ and uses two separate scan maps at an angle of 40\degree\ for
good 1/f noise reduction.  This mode provides high and relatively uniform sensitivity over an area of order
60\arcsec$\times$90\arcsec\ in the center (within 80\% of peak coverage).  Table \ref{obstable} lists the relevant parameters
for all the observations reported here.
The OBSID's for all the observations
are listed in Table \ref{aortable}.  The 1$\sigma$ noise levels for the combined OBSID's for
each object were expected to be $\sim$1 mJy at 70\micron\ and 2.3 mJy at 160\micron\ according to the
{\it Herschel} observation-planning tool HSPOT which is based on standard aperture photometry
estimates.  In fact, we achieved 10--20\%  better noise levels at 70\micron\
probably because of our use of psf-fitting photometry instead of aperture photometry.
The uncertainties at 160\micron\ were almost entirely set by the level of diffuse emission present in the images, which
differed widely between objects.

The data were first processed with the Herschel Interactive Processing Environment, HIPE version 7.3, with a script
that utilizes standard high-pass filtering for point-source observations, and which processes the
two separate OBSID's for each object together.  
The portions of scan legs where the telescope was stopped or not
scanning at a uniform velocity were not used for this processing.  The output of this script consists
of {\it fits} files of the image, coverage, and
uncertainty.  The uncertainty images are not yet reliable at this stage in the {\it Herschel} mission, so we estimated uncertainties
in additional ways described below.  

The next stage of data reduction utilized the psf-fitting photometry tool, {\it c2dphot}, originally developed by the
{\it c2d} {\it Spitzer} Legacy Team \citep{harvey06,evans07}, and based on the earlier DOPHOT tool \citep{dophot}.   
This tool can be used in various modes,
including finding peaks above the background and fitting a psf to the local maxima, or fitting a psf to any fixed
position in the image. This latter mode is especially useful for estimating the noise and determining upper limits.  The noise levels
quoted for our sources were determined this way by fitting a grid of nine positions on and around the BD position
within the high-coverage area.  Upper limits 
were also estimated this way as well as by inserting artificial stars into the images, particularly for the 160\micron\
data. 
We also computed aperture fluxes for comparison with the psf-fitted values, and
those values agree well.  In addition, we have computed aperture-flux ``curves-of-growth'' to be
sure that there are no systematic problems with the psf-fit photometry.
At 160\micron\ most of the images were contaminated substantially by structured, diffuse emission that dominated
the flux uncertainties.

\section{Observational Results}\label{obsresults}

\begin{figure}
\plotfiddle{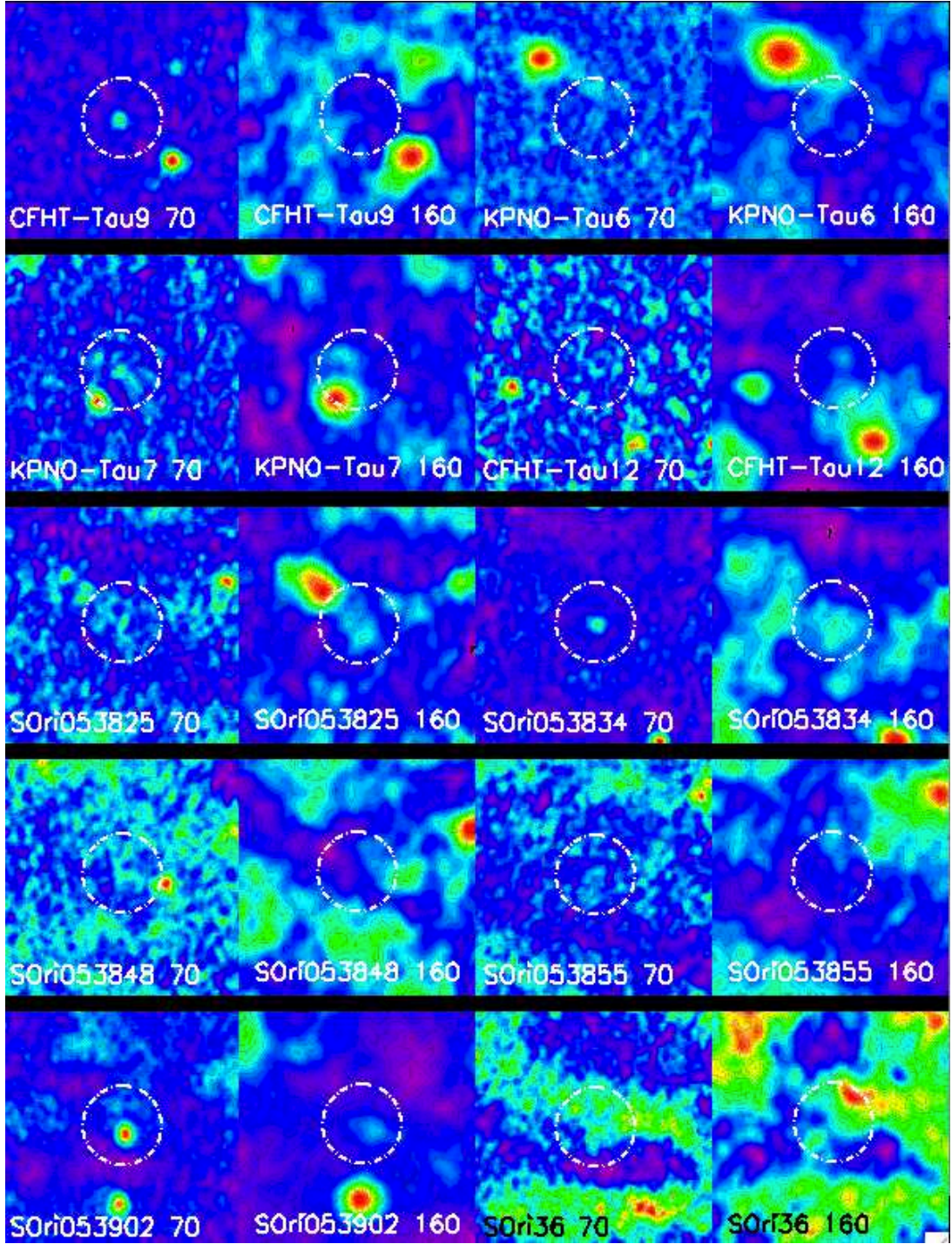}{7.5in}{0}{70}{70}{-220}{-00}
\figcaption{\label{imfig1}
Images of first 10 objects in Table \ref{aortable}. Dotted circle is 30\as\ in diameter. }
\end{figure}

\begin{figure}
\plotfiddle{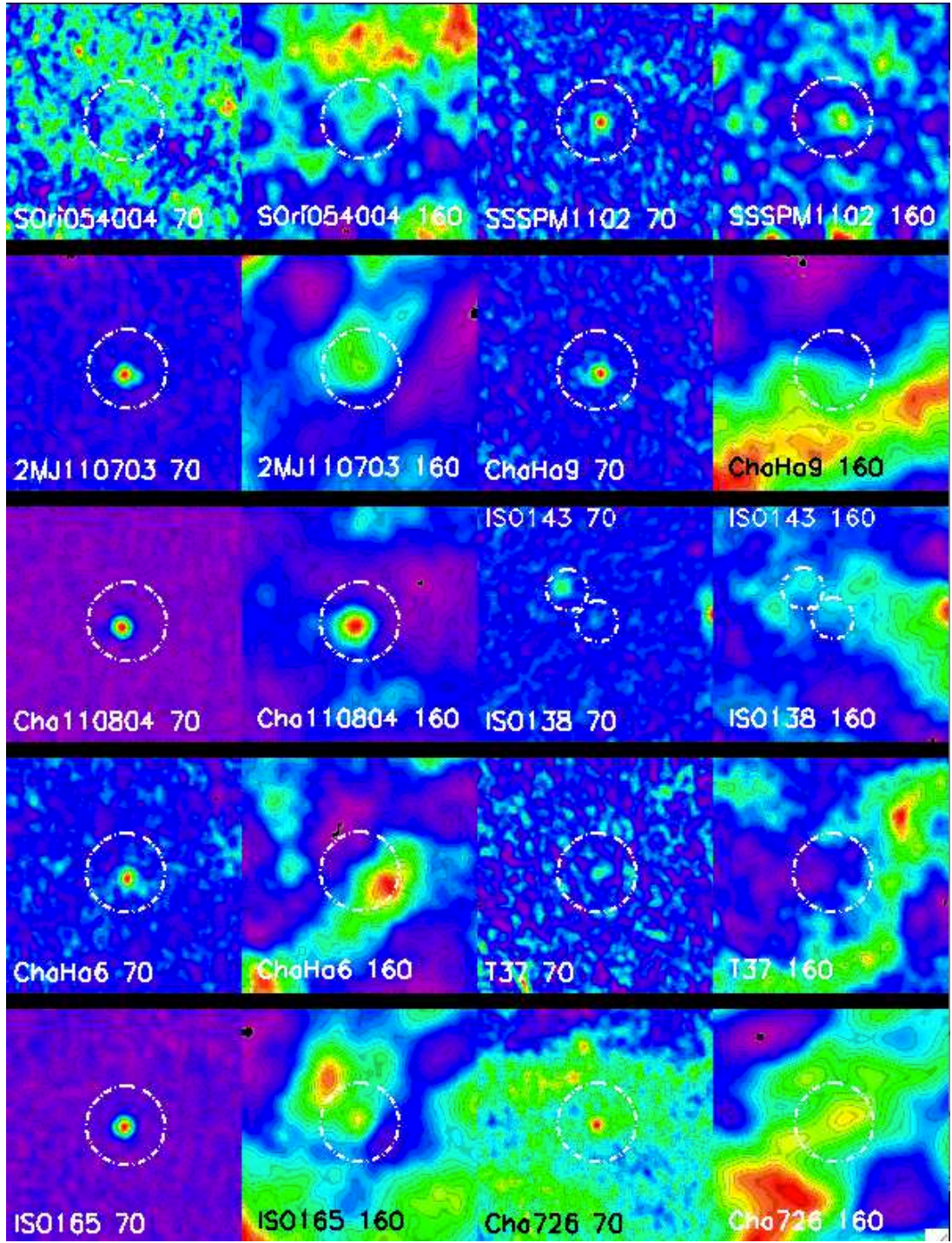}{6.5in}{0}{70}{70}{-220}{-00}
\figcaption{\label{imfig2}
Images of next 11 objects in Table \ref{aortable}.  Circle diameter for ISO138/143 is 16\as.}
\end{figure}

\begin{figure}
\plotfiddle{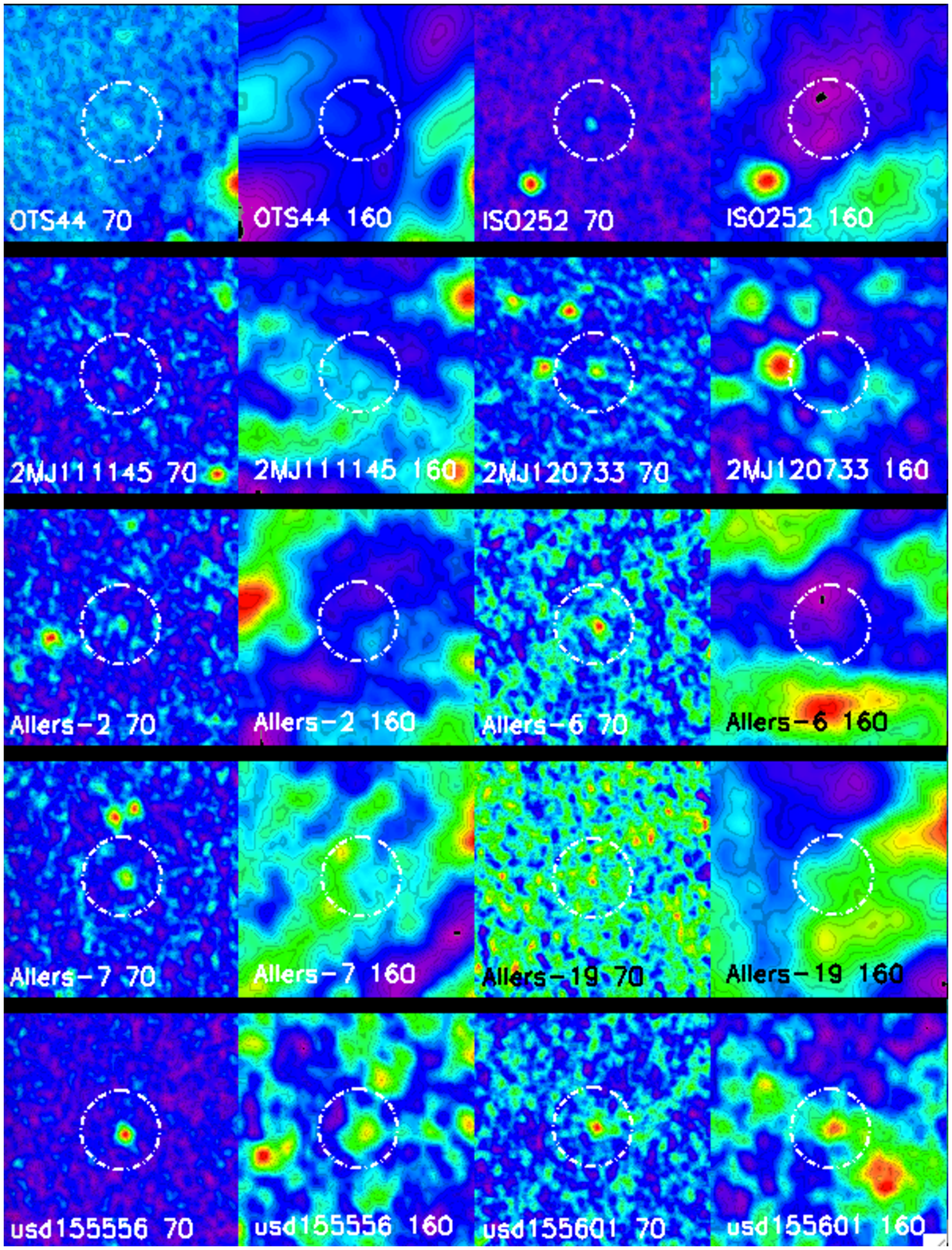}{6.5in}{0}{70}{70}{-220}{-00}
\figcaption{\label{imfig3}
Images of next 10 objects in Table \ref{aortable}. }
\end{figure}

\begin{figure}
\plotfiddle{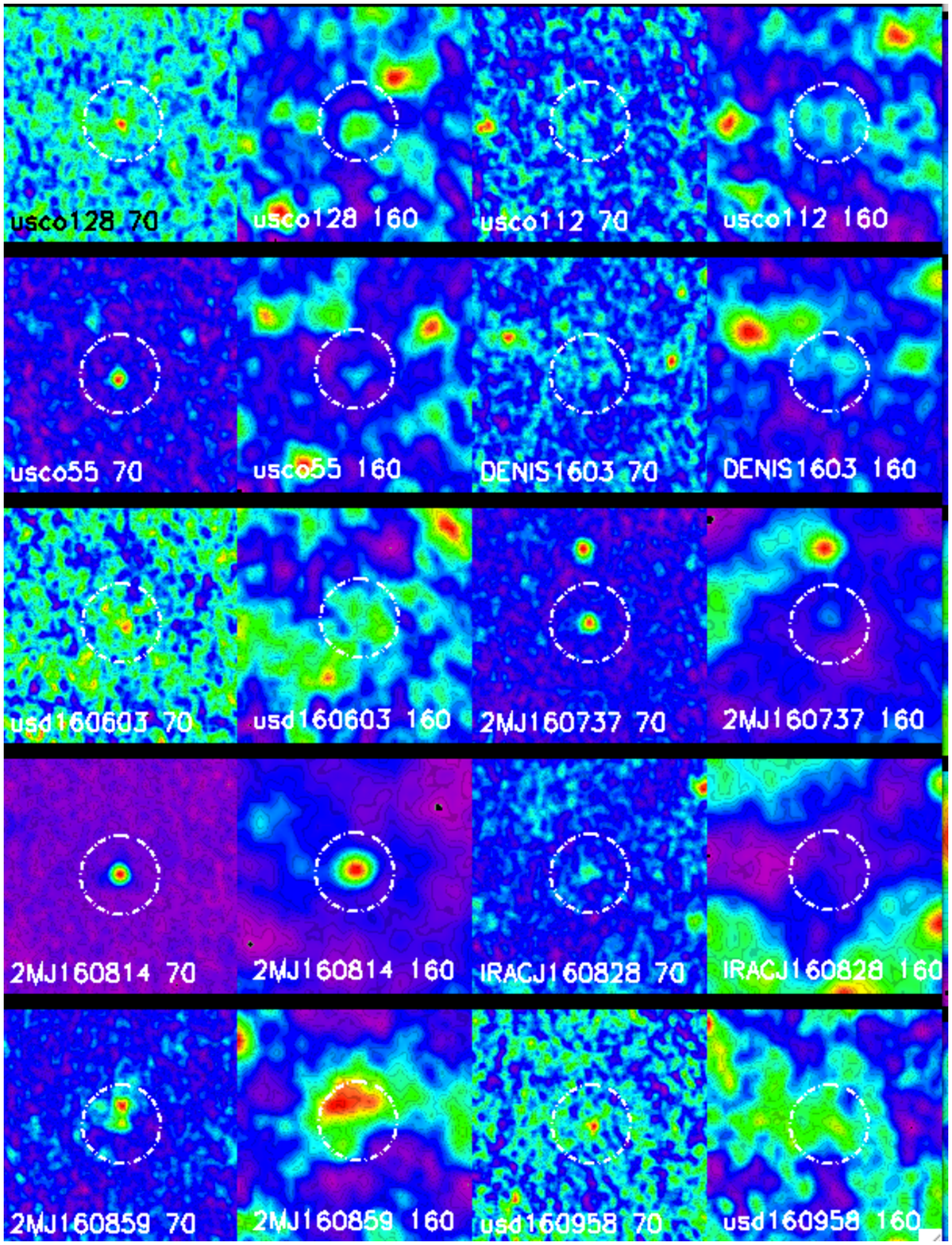}{6.5in}{0}{70}{70}{-220}{-00}
\figcaption{\label{imfig4}
Images of next 10 objects in Table \ref{aortable}.}
\end{figure}

\begin{figure}
\plotfiddle{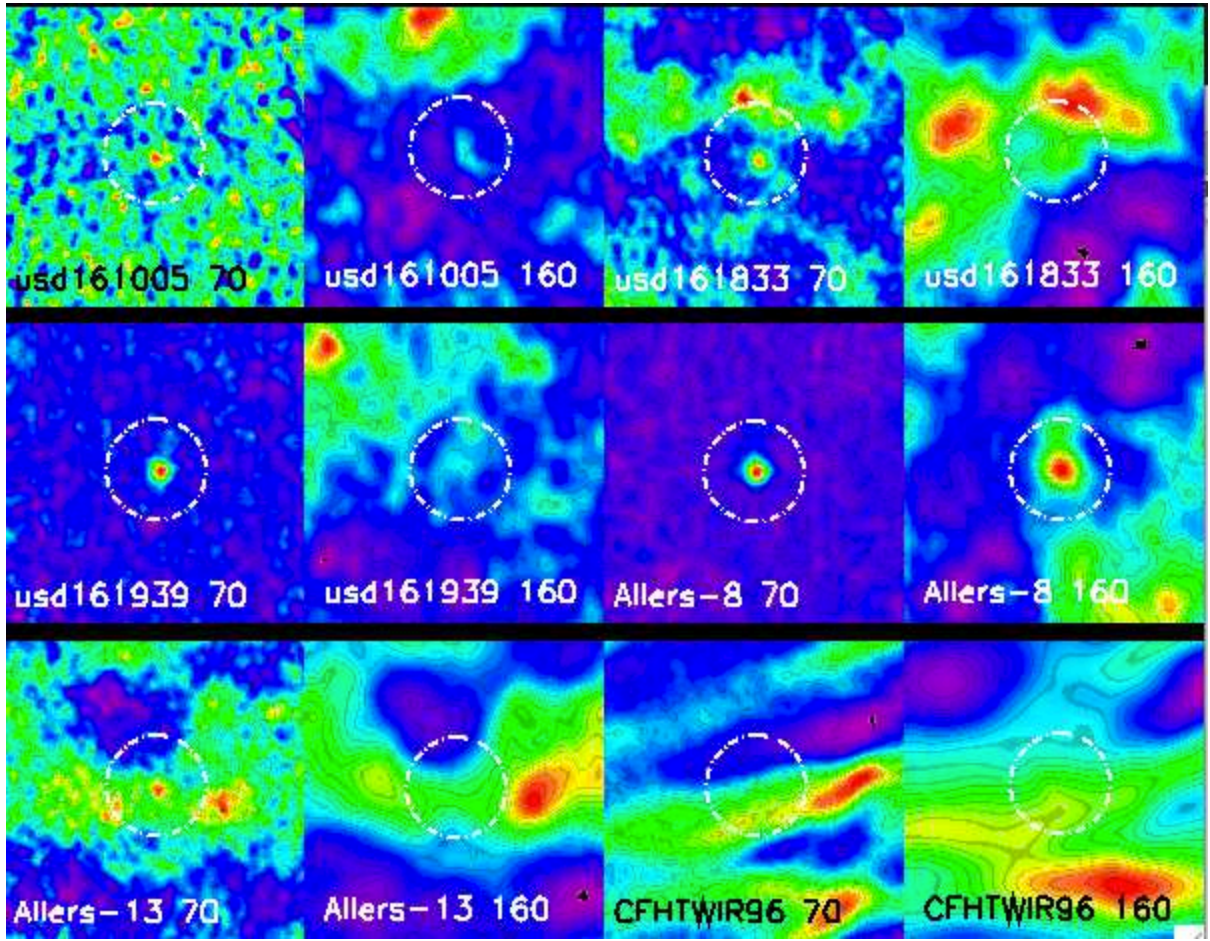}{6.5in}{-90}{60}{60}{-250}{400}
\figcaption{\label{imfig5}
Images of last 6 objects in Table \ref{aortable}.}
\end{figure}

Figures \ref{imfig1}--\ref{imfig5} (electronic version only) show images of all 47 of the observed low mass young objects.
A summary of our measured flux densities is given in Table \ref{aortable}.  We have not included the four objects
found to be extragalactic, but do list the object cha110804 described as ``possible Class 1'' by \citet{luhman08a}
since its luminosity is probably in the same range as the other objects.  The entries in Table \ref{aortable} given as
upper limits have been determined by finding the maximum brightness source that could have been present at the position
of the object without us reliably detecting it; therefore, these should be viewed as firm limits, equivalent to $\sim 3\sigma$.  For objects where
an examination of the image showed a possible detection, we report the formal flux level derived from the psf-fit 
and the uncertainty determined as described above using multiple psf-fit positions in blank parts of the image.  

\begin{figure}
\plotfiddle{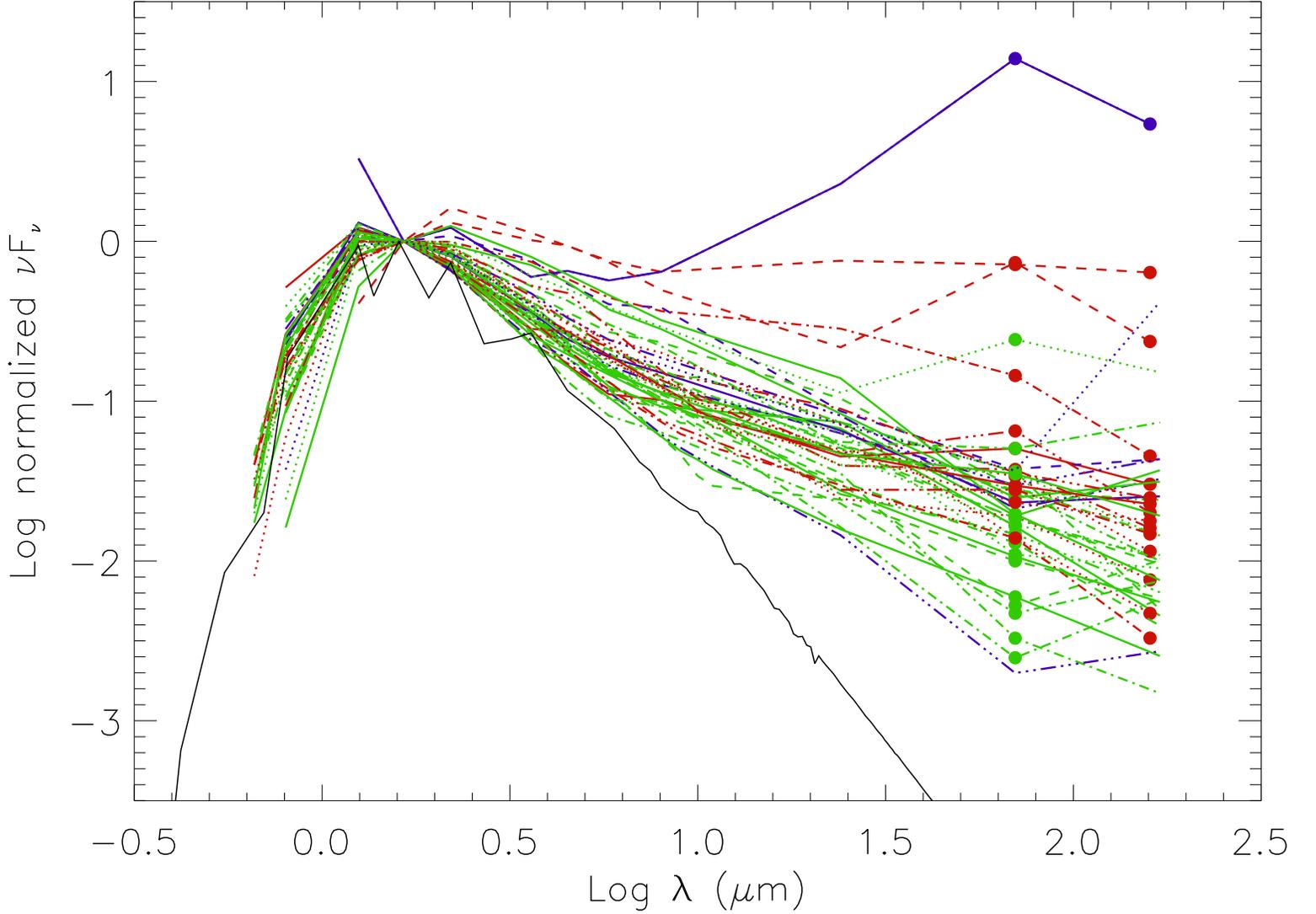}{5.0in}{90}{80}{80}{320}{10}
\figcaption{\label{sedall}
Overplotted SED's normalized to 1.6\micron. Objects with S/N $>$ 2 at both 70 and 160\micron\ are shown
with fluxes plotted as red circles and red SED's.  Objects with S/N $>2$ only at 70\micron\ are shown plotted in
green, and remaining objects without reliable detections at either wavelength are in blue.  Plotted fluxes without
circles are only upper limits or low S/N.  The solid black curve shows a typical SED of a bare BD photosphere
(SSSPM110209 from Paper I).  The possible Class I object, Cha110804, is the solid blue line with the highest
values of normalized 70 and 160\micron\ fluxes; note flux values for  his object are upper limits shortward of 3.6\micron\
as measured from the 2MASS survey images.}
\end{figure}

\begin{figure}
\plotfiddle{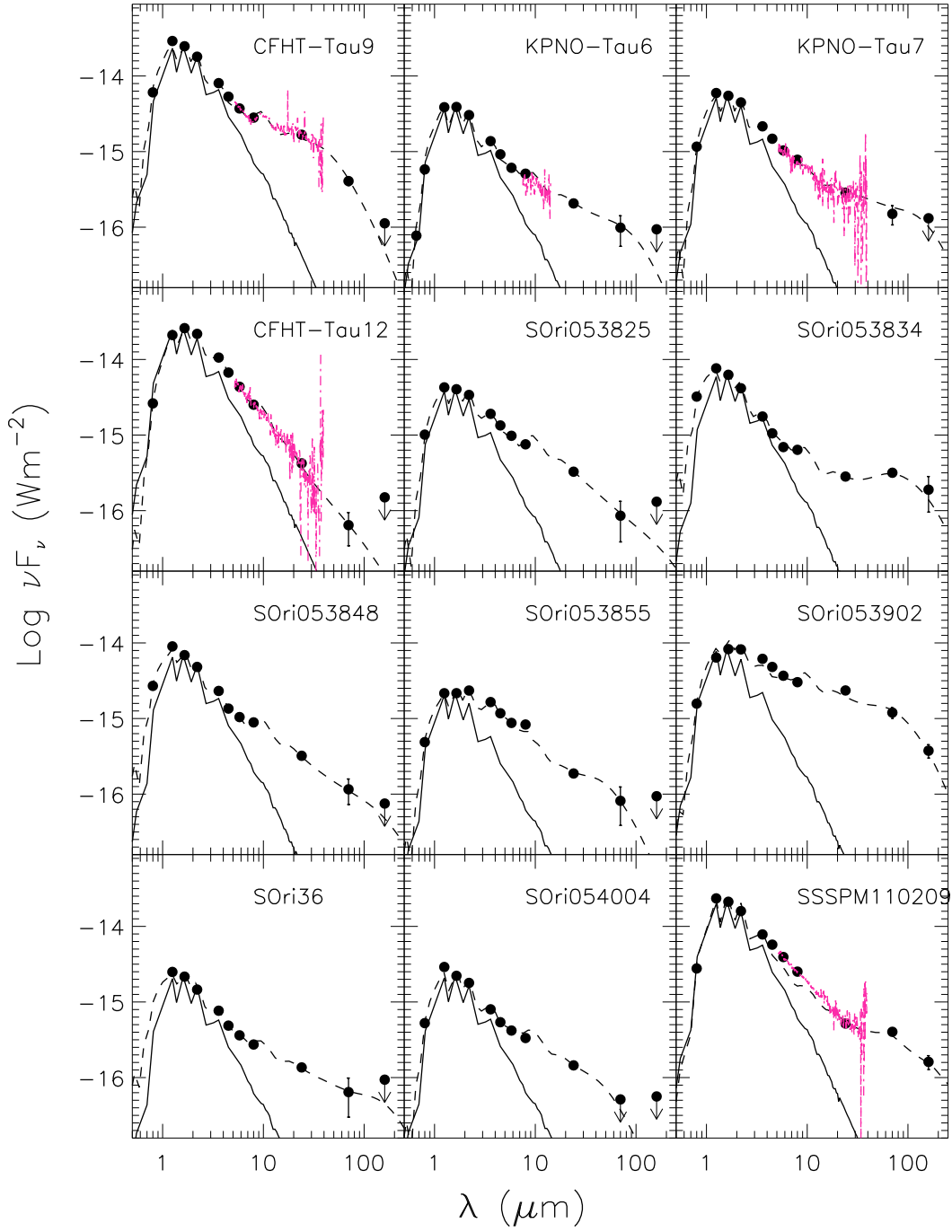}{6.5in}{0}{70}{70}{-220}{-00}
\figcaption{\label{sedind1}
SED's of the first 12 Class II objects in Table \ref{aortable}.  Representative model fits are shown with
dashed lines, and Spitzer IRS data are shown in red.  A typical bare photosphere is shown in solid, normalized
to the observed 1.6\micron\ flux.}
\end{figure}

\begin{figure}
\plotfiddle{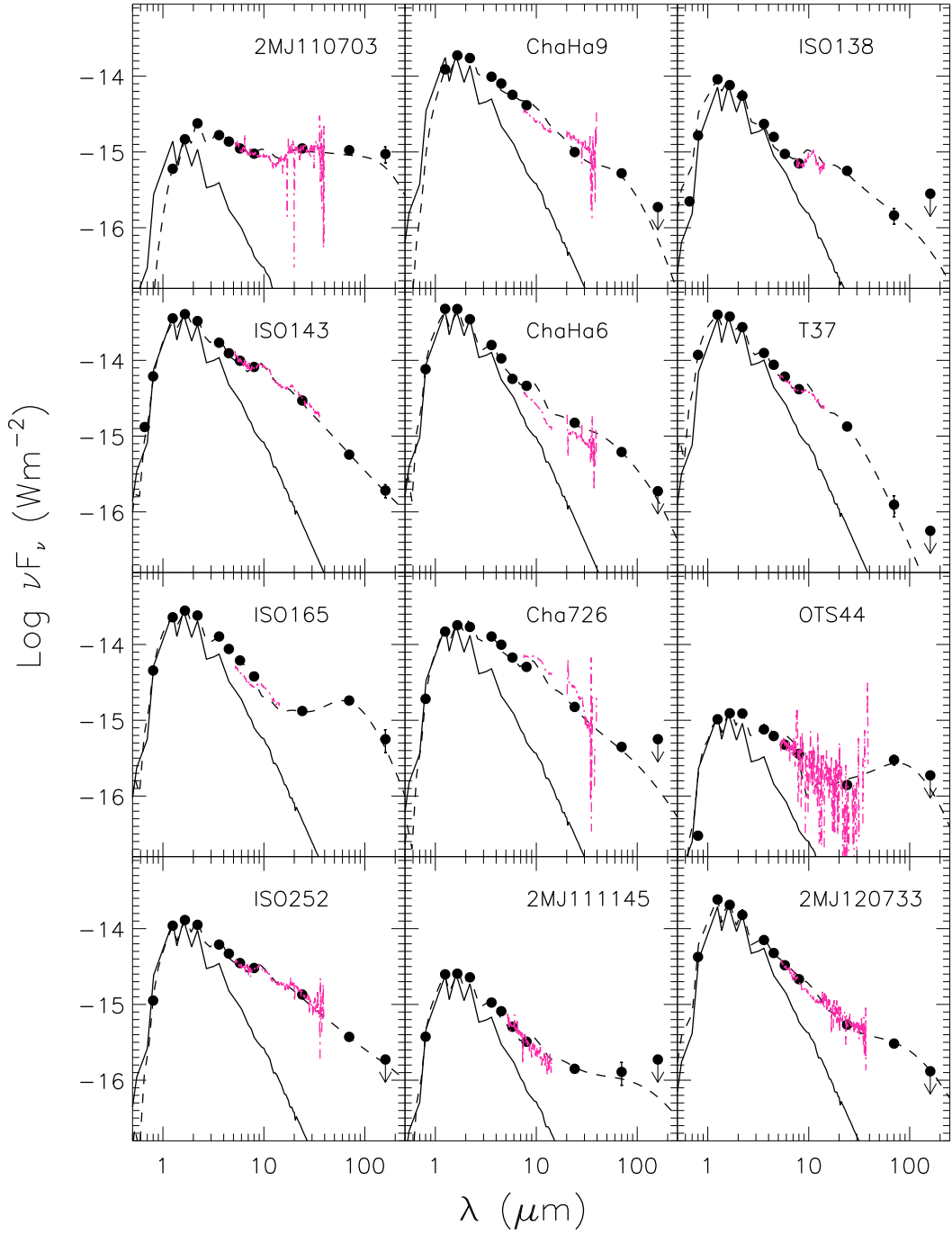}{6.5in}{0}{70}{70}{-220}{-00}
\figcaption{\label{sedind2}
SED's of the second 12 Class II objects in Table \ref{aortable}.}
\end{figure}

\begin{figure}
\plotfiddle{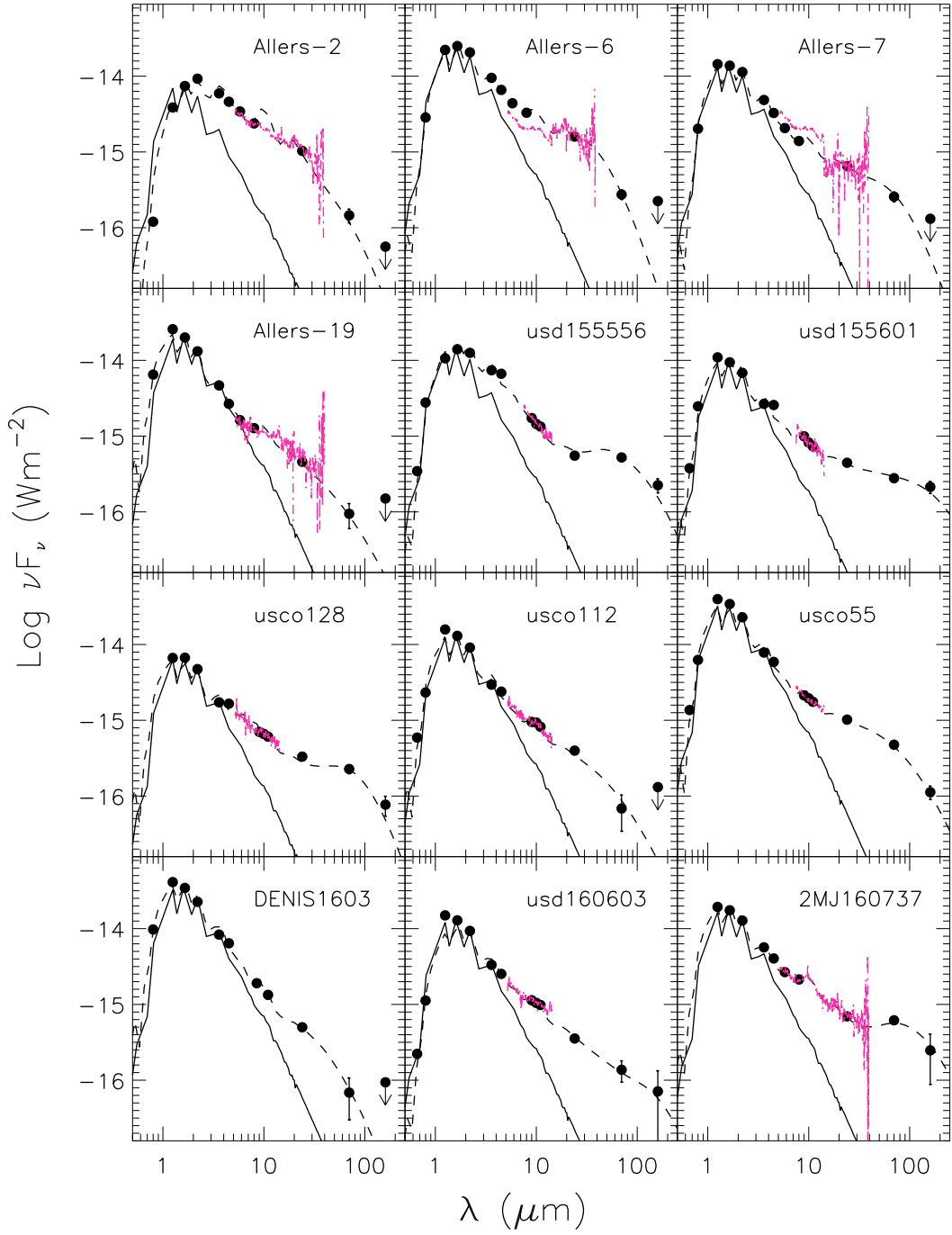}{6.5in}{0}{70}{70}{-220}{-00}
\figcaption{\label{sedind3}
SED's of the third 12 Class II objects in Table \ref{aortable}.}
\end{figure}

\begin{figure}
\plotfiddle{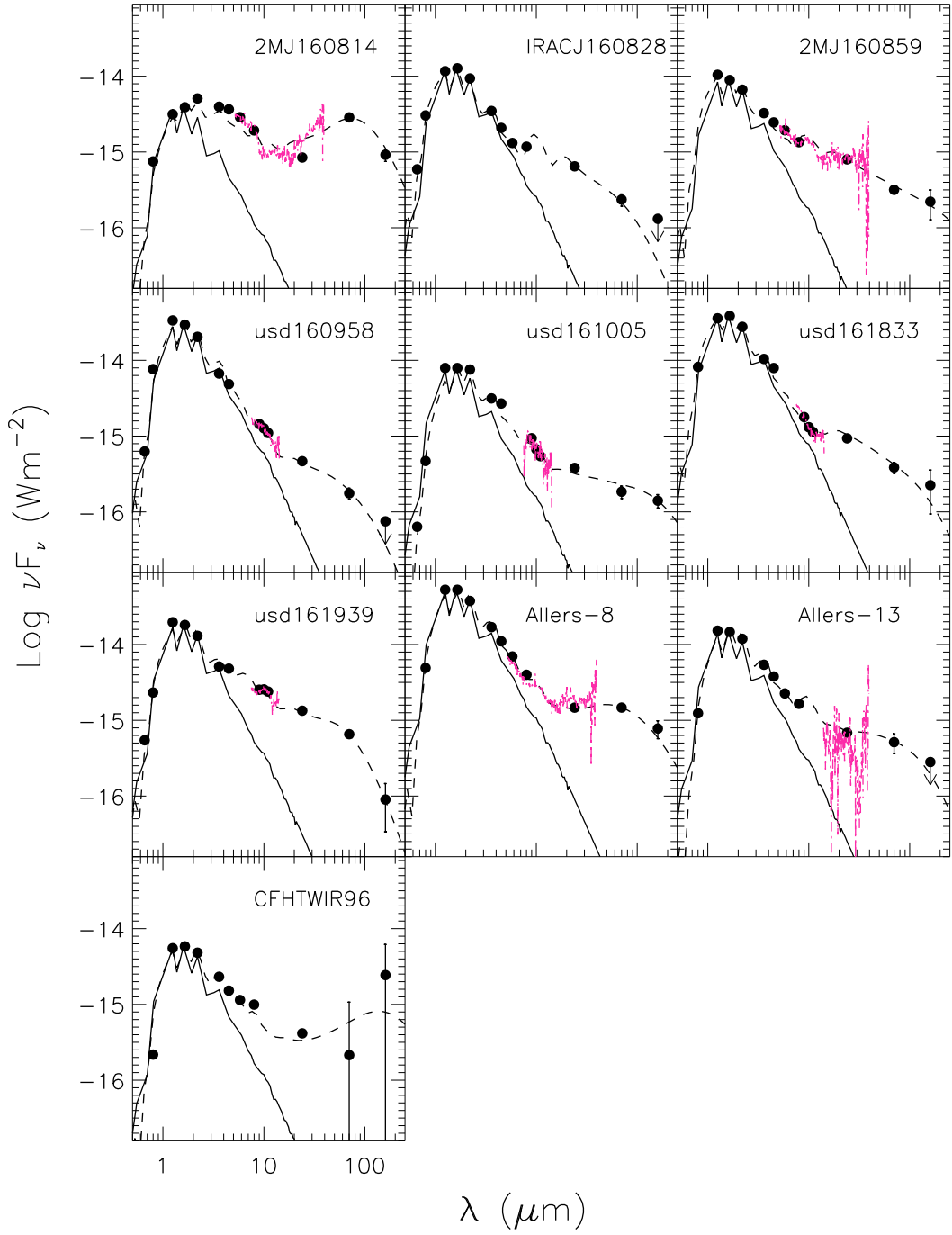}{6.5in}{0}{70}{70}{-220}{-00}
\figcaption{\label{sedind4}
SED's of the last 10 Class II objects in Table \ref{aortable}.}
\end{figure}

All of the objects in Table \ref{aortable}
have been observed over a wide range of wavelengths by some or all of the 2MASS \citep{2mass} and DENIS surveys, the recent WISE survey
\citep{wise}, numerous {\it Spitzer} programs using IRAC and MIPS, as well as 36 with the IRS instrument on {\it Spitzer}.
This enables us to present very complete SED's for all the sources, and as discussed below, enables model fitting
to derive limits on physical parameters of the circumstellar disks.
We show these in two forms.  Figure \ref{sedall} shows the SED's of all the
objects superposed and normalized to the 1.6\micron\ flux density (or upper limit in the case of cha110804).  Figures \ref{sedind1} --
\ref{sedind4} show the individual SED's for all the objects except cha110804.

\begin{figure}
\plotfiddle{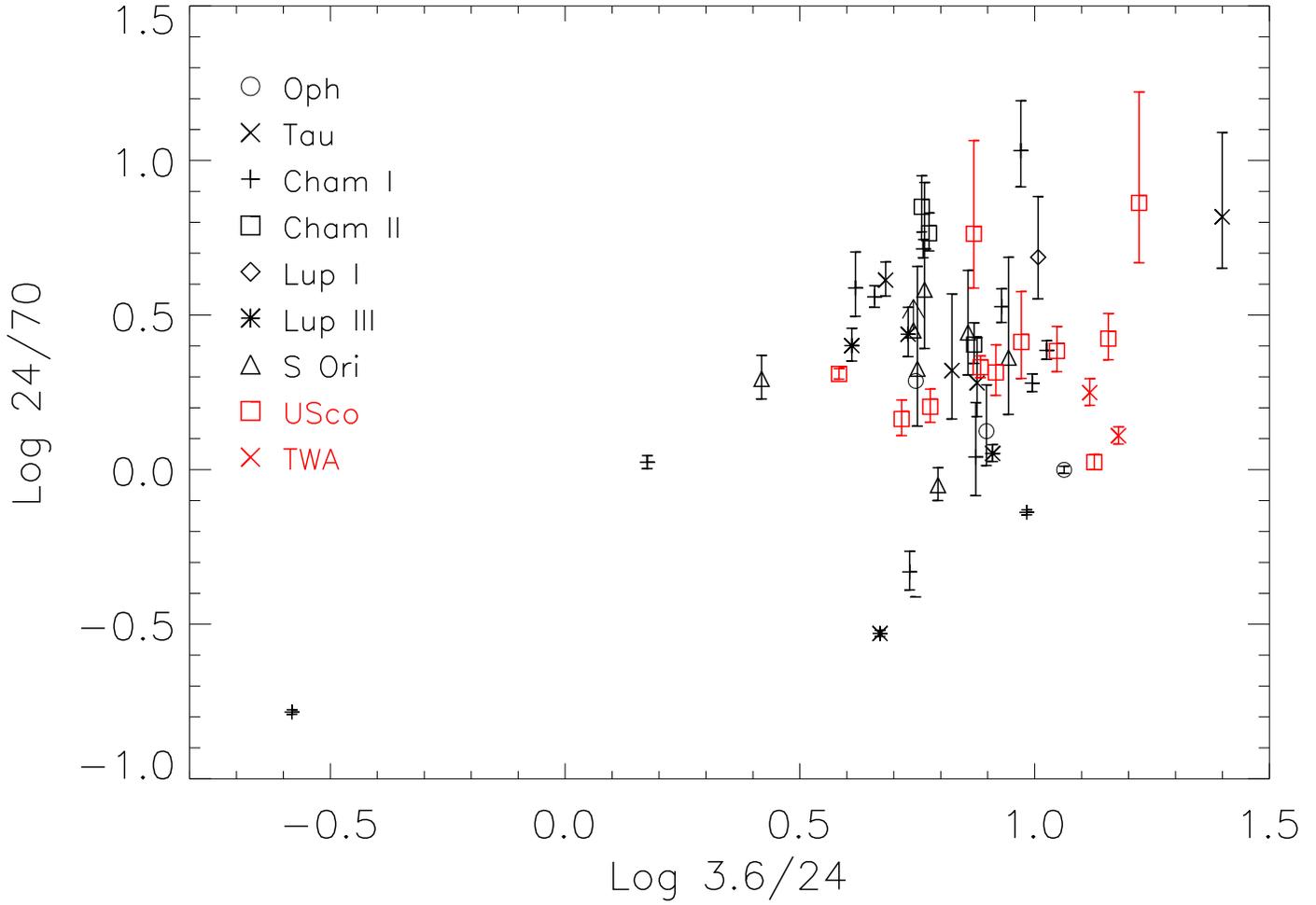}{6.5in}{90}{80}{80}{290}{70}
\figcaption{\label{color}
Color-color diagram showing far-ir color ($\nu F_{\nu}$ 24/70) versus mid-ir color ($\nu F_{\nu}$ 3.6/24) as a function of
star-forming cloud in which the object is found.  There is no clear trend with region, suggesting that
there is no obvious dependence of outer disk structure/mass versus age.}
\end{figure}

Figure \ref{sedall} shows that the observed SED's between 1 and 100\micron\ differ considerably, and it suggests also
that there appears to be a relatively continuous distribution of slopes in the mid- to far-infrared.  We can
also illustrate this conclusion with the color-color diagram in Figure \ref{color} that shows a plot of
the ratio of $\nu$F$_{\nu}$ at 24\micron\ versus 70\micron\ compared to the comparable ratio at 3.6\micron\ versus 24\micron.
This also shows a relatively continuous distribution of colors, although 90\% of the objects have $\nu$F$_{\nu}$(24\micron)
$\ge$  $\nu$F$_{\nu}$(70\micron).
The very red object cha110804 is clearly beyond the
end of the distribution of the other objects that are believed to have circumstellar disks but not substantial
envelopes.  

One question we can consider with just this simplied analysis of the SED's is whether there is any evidence
for differences in disks with age or environment.
The symbols for each object in Figure \ref{color} have been coded to denote the host cloud
in which they are found.  Within the relatively small number statistics for some of the regions, there is no
clear trend in far-ir excess as a function of the location/age of the cloud in which each
object is found.  The same conclusions holds if we restrict the sample to a subset of objects with well-determined
spectral types in the M5--6 range which has a reasonable number of sources.

\section{Radiative Transfer Modeling}\label{model}

In Paper I we showed results from modeling the disks around the first three objects observed in this
program, 2MASS120733, ISO138, and SSSPM110209.  Despite the degeneracy inherent in such modeling,
we found that for those three cases with excellent SED coverage we were able to constrain the
circum-object disk masses and geometry to some degree.  
For this current study we performed radiative transfer modeling of the complete set of objects,
though the results are probably most useful for those with the most
complete SED coverage.  We have not tried extensively to fit the Spitzer IRS
data in the area of the silicate feature which is generally viewed as arising in the disk
atmospheres, e.g. \citet{olof09}, and we have assumed a single, uniform grain composition.  

We used the two radiative transfer codes discussed already in Paper I, MC3D described by \citet{wolf99, wolf03}
and MCFOST described by \citet{pinte06, pinte09}.  Both codes are three-dimensional, radiative transfer
codes using the Monte-Carlo method.  NextGen stellar atmosphere models \citep{nxtgen1, nxtgen2} were used for
the stellar input to the disk models with the photospheric temperature set by the assumed spectral type
where available.
With each code we modeled each SED with a parametric disk.
We use a dust density distribution with a Gaussian vertical profile
$\rho(r,z)=\rho_0(r)\,\exp(-z^2/2\,h^2(r))$, valid for a
vertically isothermal, hydrostatic, non self-gravitating disk.
We use power-law distributions for the dust surface density
$\Sigma(r) = \Sigma_0\,(r/r_0)^{-p}$  and the scale height $
h(r) = h_0\, (r/r_0)^{\gamma}$ where $r$ is the radial coordinate
in the equatorial plane.  The parameter $h_0$ is the scale height at a fiducial radius chosen to be
$r_0 = $ 100 AU for easier comparison with previous BD and T Tauri disk models. The scale height parameters, $\gamma$ and $h_0$, 
are free parameters in the modeling and not calculated from assumptions of hydrostatic equilibrium.  The disk extends from an inner radius  $r_{\rm
in}$  to an outer limit radius $r_{\rm out}$. The central star
is represented by a sphere radiating uniformly.  

We consider homogeneous spherical grains and we use the
dielectric constants of astro-silicates \citep{draine03}. The
differential grain size distribution is given by $\mathrm{d}n(a)
\propto a^{-3.5}\,\mathrm{d}a $ with grain sizes between
$a_{\mathrm{min}} = 0.03 \mu$m and  $a_{\mathrm{max}} = 1 \mu$m or 1mm.
The mean grain density is 3.5 g cm$^{-3}$.  Dust extinction and scattering opacities, scattering
phase functions and Mueller matrices are calculated using Mie
theory. 
The grain properties are taken to be independent of position within the disk.
All disk masses discussed below are described as ``Total'' disk mass
of gas and dust together. 
We make no assumptions about the gas distribution in the disk, but
we assume a ratio of 100 for gas mass relative to dust mass for comparison with
other disk mass estimates.  Of course
the SED modeling is only sensitive to the dust mass, and even in that case only to dust up to
some nominal size of a few times the longest observed wavelength.
In principle, the two radiative transfer codes should give the same result, but because of differences in the exact model
grid, there were sometimes modest differences.  For example, the MCFOST grid in mass was spaced by 0.5
in $log_{10}(Mass)$ from $10^{-6} - 10^{-2}$ \msun\ (total disk mass), while the MC3D grid was spaced by 1.0 in $log_{10}(Mass)$ from
$5 \times 10^{-8} - 5 \times 10^{-4}$ \msun.

\begin{figure}
\plotfiddle{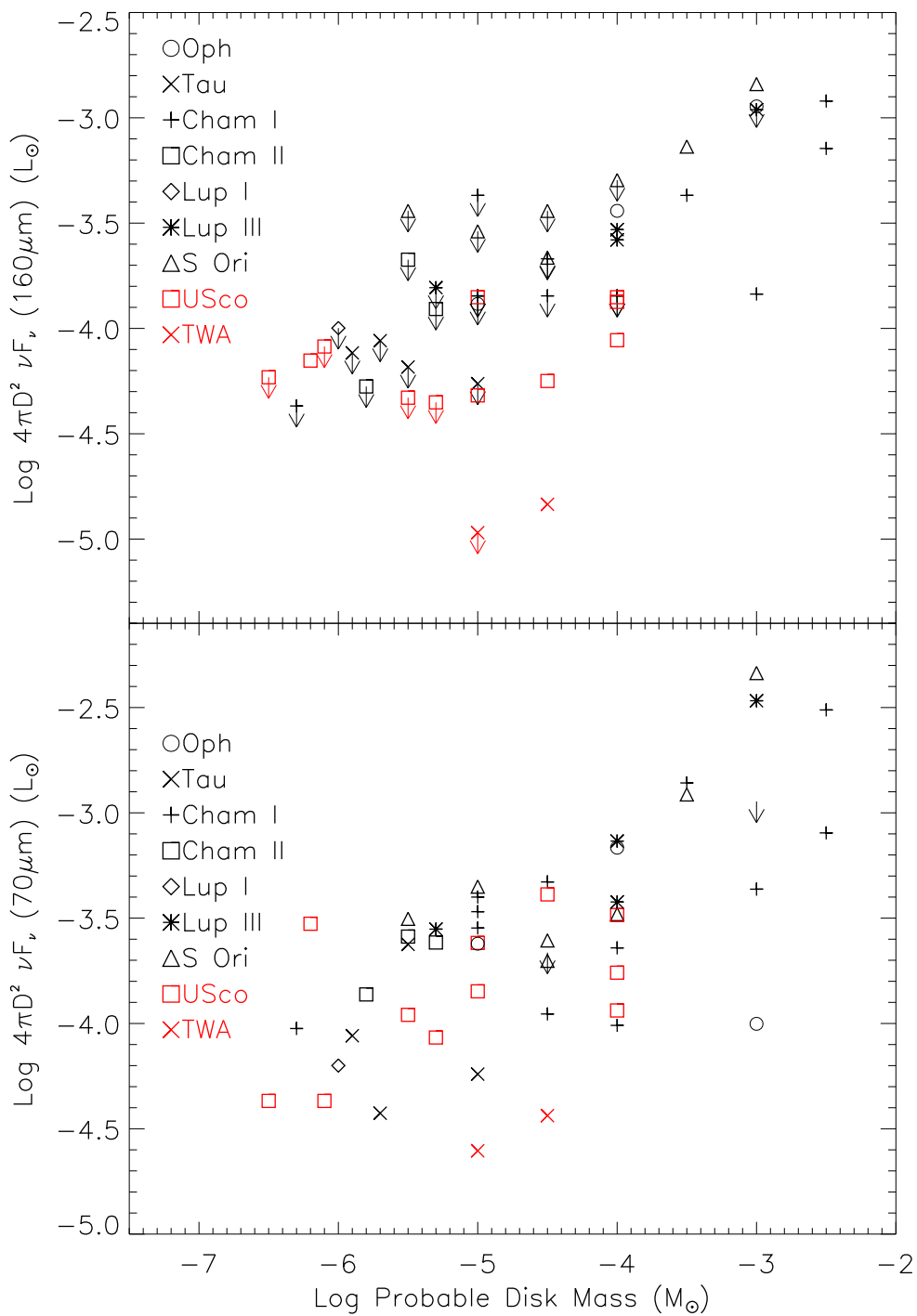}{7.0in}{00}{70}{70}{-190}{00}
\figcaption{\label{mass70160}
Comparison of the 70 and 160\micron\ luminosities with the model-derived most probable disk
masses illustrating the
utility of the longest PACS wavelength for estimating disk masses.}
\end{figure}

As discussed in Paper I, the choice of outer radius for the disks makes essentially
no difference to the model SED in the spectral range of our study because most of the far-ir emission comes from dust at
radii inside of 20 AU; so this was left fixed in
all the models at 100 AU.  The choice of outer radius also makes only a small difference in the probable values of most
other disk parameters.  This is shown both in Paper I by a comparison of the disk parameters for 2MASS120733 for outer radii of
of 75 versus 25 AU, and by the modeling described by \citet{bouy08} for 2MASS04442713.  For example, the total disk mass
is driven largely by the total far-infrared/submm flux (see also discussion below and Figure \ref{mass70160}).

Because of the 
remaining degeneracy in the model-fitting despite our extended SED coverage, we present the results
largely from the Bayesian probability analysis for various disk parameters.  In other words, at this point
in the model-fitting it is not appropriate to choose a ``best-fit'' model or model parameter(s) but simply
to state the most probable range for disk parameters.  For example, sometimes the model with the absolutely
lowest chi-squared implies a parameter value somewhat offset from the peak of its probability distribution,
since the probability distribution for each parameter is marginalized over all possible values of the
remaining parameters.  This can lead to noticeable differences between some of the disk parameters 
for two models that fit the SED nearly equally well.

\begin{figure}
\plotfiddle{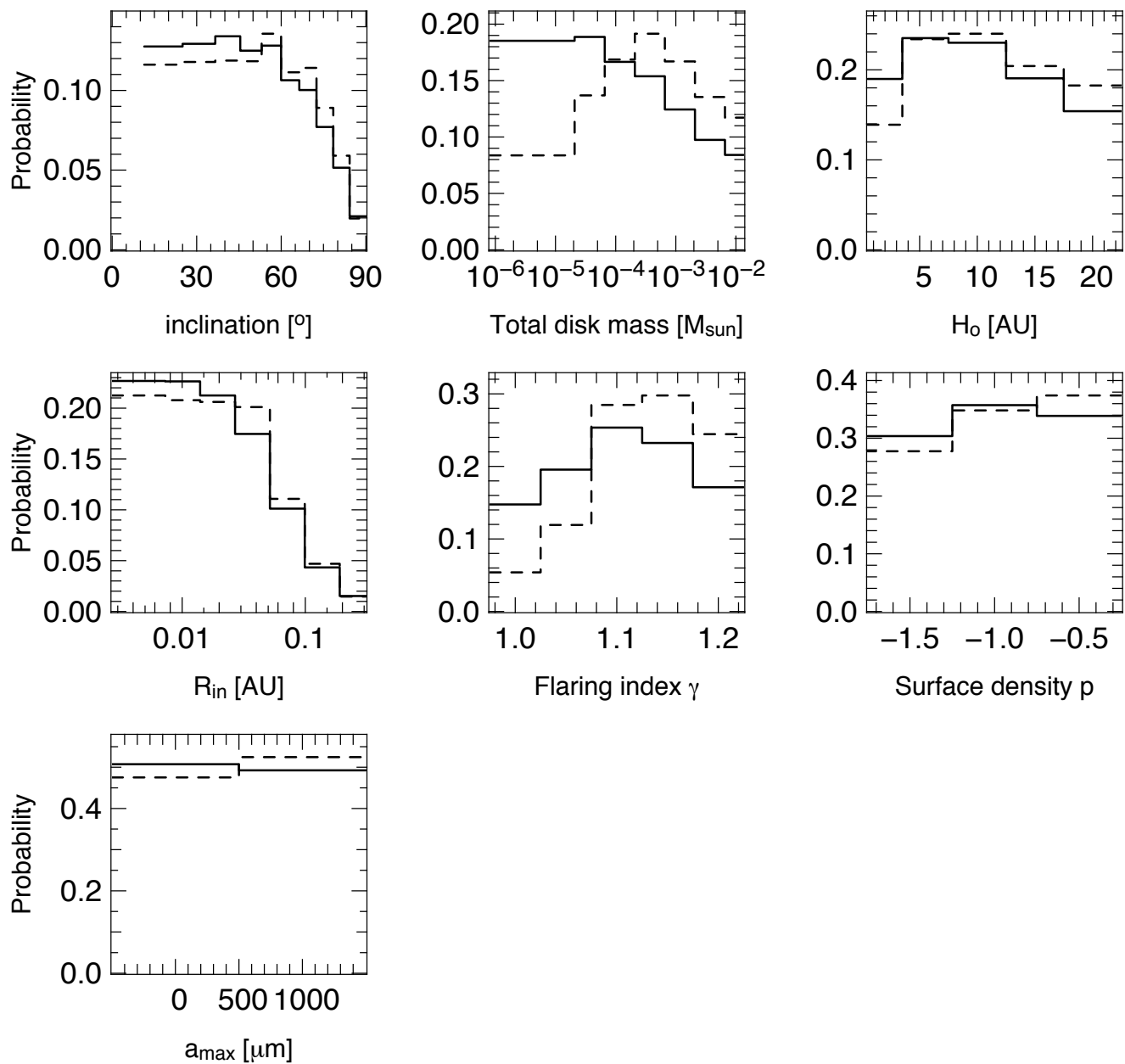}{6.5in}{0}{85}{85}{-250}{-50}
\figcaption{\label{histoplots}
Histograms of the total of all probability distributions for the disk parameters illustrating the likely
range of these parameters in the total sample.  The solid line is for the entire sample and the dashed
line is for just the objects detected at 160\micron.
}
\end{figure}

\begin{figure}
\plotfiddle{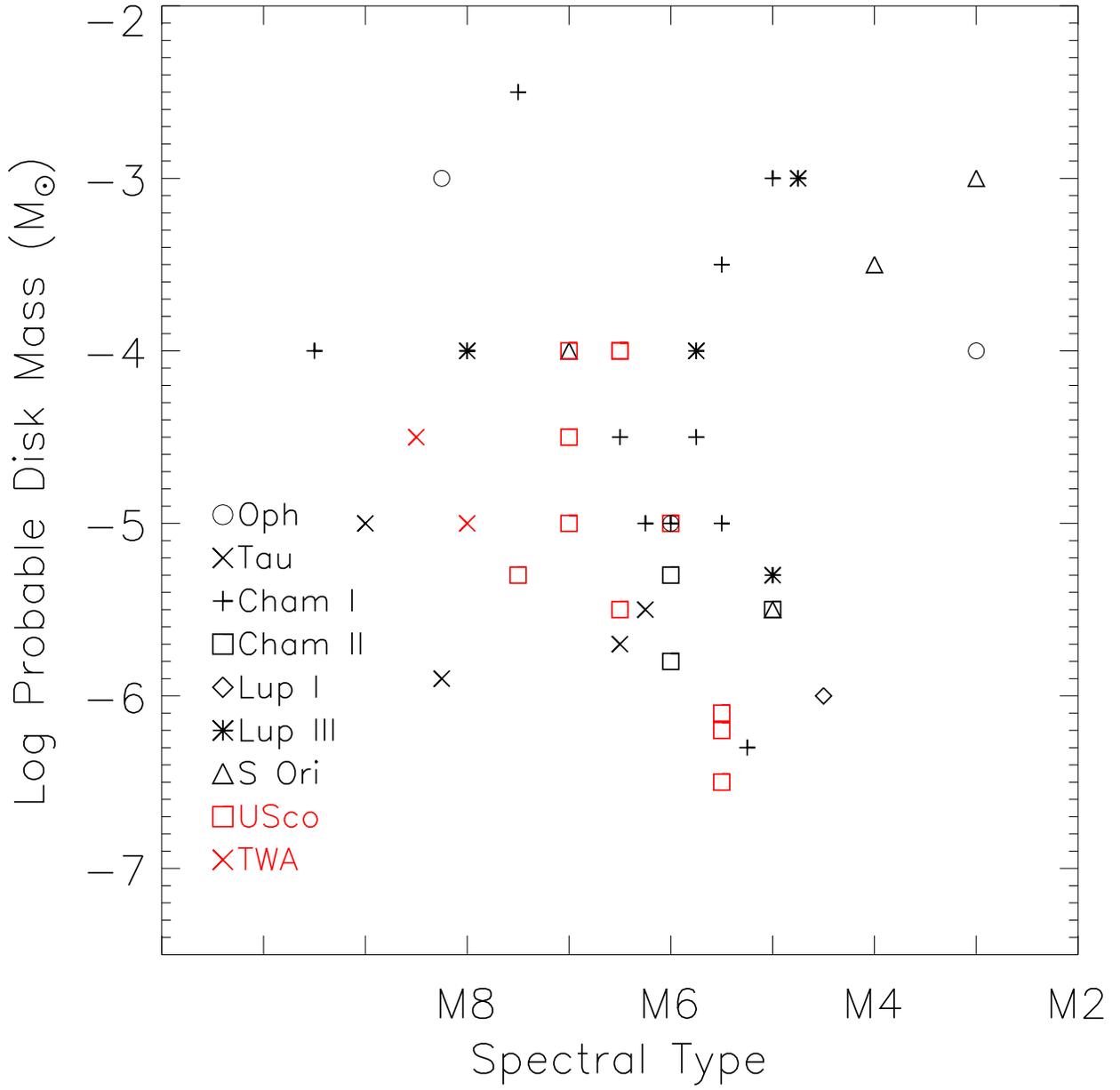}{6.5in}{90}{90}{90}{390}{00}
\figcaption{\label{massspt}
Most probable disk mass versus spectral type for the 43 objects with published spectral types.}
\end{figure}

The overall results of the modeling are displayed in several figures.  Figures \ref{sedind1}--\ref{sedind4} show example
model SED's from the set that produces reasonable fits for each object.  Figure \ref{histoplots} shows histograms of
the total of the probability distributions for all the program objects as well as the subset detected at 160\micron\
to show the likely average distribution
of these parameters.
Figure \ref{massspt} shows the distribution of the most probable disk masses as a function of spectral type for the 43 objects with
published spectral types, and finally Figure \ref{mass70160} shows the distribution of probable disk masses versus
the 70 and 160\micron\ luminosities for each object.  We discuss these results below.
More refined modeling of individual sources will be pursued in a subsequent study where we will report
individual disk parameters.

In Paper I we discussed the comparison between our model results for the first three objects and several existing
models in the literature for those objects.
To our knowledge only
two additional objects from our sample, CFHT Tau 9 and OTS 44, have a published model. In a recent study \citet{riaz12a} 
investigated the radial dependence of grain
composition in brown dwarfs relative to T Tauri stars.   Since they used a geometrical model of the disk similar to those in our study, it
is possible to compare the resulting model fits directly.  In particular, for CFHT Tau 9 without the additional spectral
coverage of {\it Herschel}, they found an inner disk radius equal to
the sublimation radius, a disk inclination of order 30--40\degree, and a flaring index of 1.1 provide a good fit to the overall
SED.  These values are essentially identical to the probability peaks in our fits for this object.
\citet{luhman05} have modeled the disk around OTS44, but their model was based on different disk parameters making it
difficult to perform a detailed comparison.
Finally, \citet{riaz12b} have recently reported a large submm flux in the {\it Herschel} SPIRE bands from 2M1207 that would
imply a much larger disk mass than found from our modeling, but  have since retracted that conclusion \citep{riaz12c}.

\section{Discussion}\label{disc}

\subsection{Overview of Modeling Results}

Despite the inherent degeneracies in SED modeling without angular size information, there are
several conclusions that appear relatively firm as illustrated in the figures described above.
The majority of the disks are best modeled with scale heights $h_o$ at the fiducial radius of 100 AU of order 5 -- 20 AU
and radial scale height dependences $\gamma$ in the range 1.1 -- 1.2.  Their inner radii are likely
close to the dust sublimation radii which are typically of order $5 \times 10^{-3}$ AU for the faintest objects
up to $\sim 1.5 \times 10^{-2}$ AU for the most luminous. Their surface density gradients are best 
fit with a relatively shallow slope of $p \sim 0.5-1.0$ ($\Sigma \propto r^{-p}$) although this
value is not well constrained.  There is clearly a very wide range in
likely masses with a median value of order $3 \times 10^{-5}$ \msun\ (0.03 $M_{Jup}$), a value which is
roughly the same for both the objects with spectral types later than M6 and earlier than M6.  Unless these masses are
severely underestimated due to grain growth to sizes well beyond a value to which this study is
sensitive, this suggests that future giant planet formation would be very difficult in these disks.
If we assume that spectral type is a rough proxy for brown dwarf mass, then Figure \ref{massspt} suggests 
that the ratio of disk mass to sub-stellar mass covers quite a wide range, of order a few $\times\ 10^{-5}$
up to $\sim 10^{-2}$.

\begin{figure}
\plotfiddle{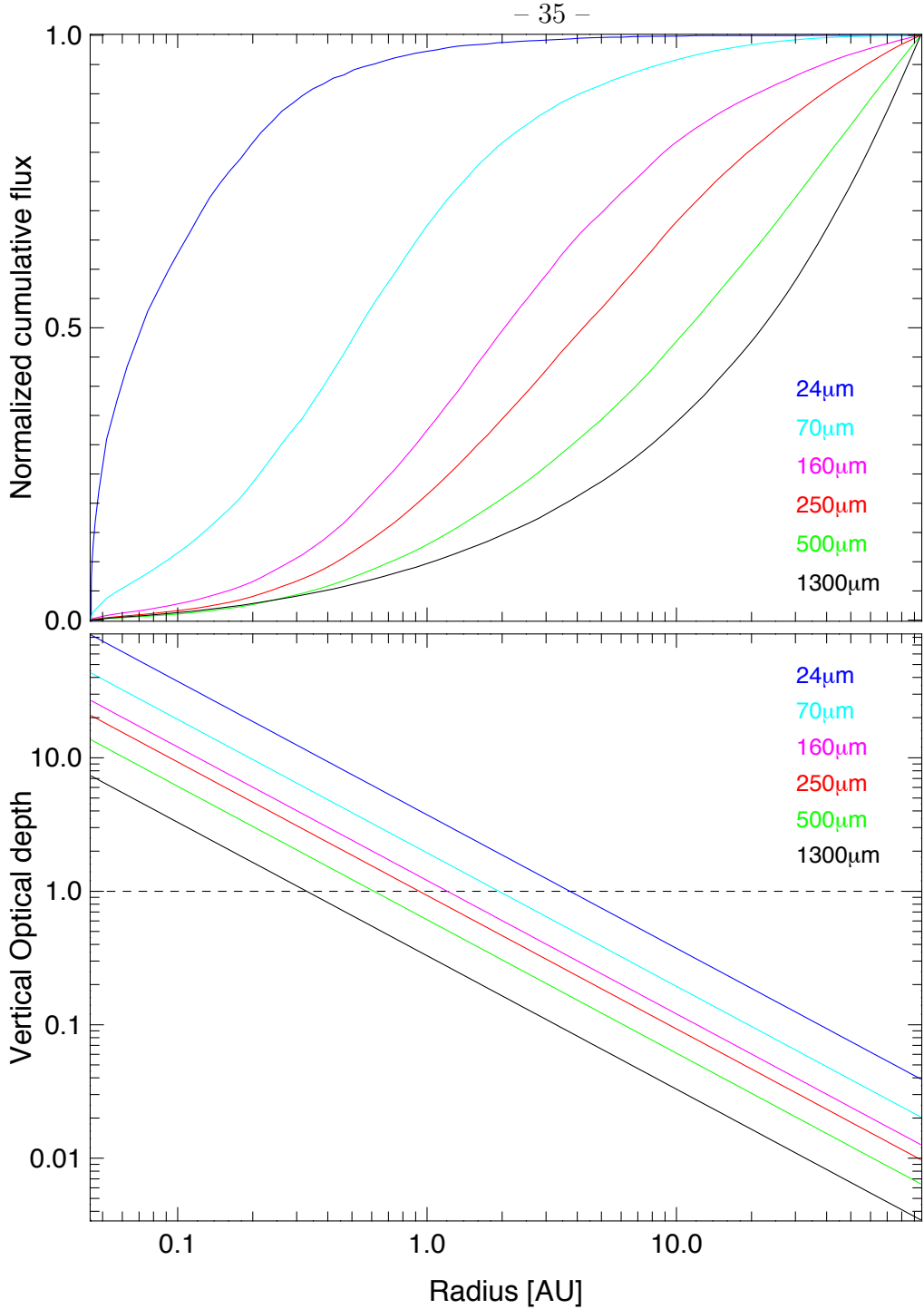}{7.0in}{0}{70}{70}{-250}{-00}
\figcaption{\label{taufig}
Upper panel -- plot of cumulative emitted flux as a function of radius at several wavelengths for
a typical disk model with $M_{disk} = 2 \times 10^{-4}$ \msun, $R_{in} = 0.05$ AU, $a_{max} =$ 1mm, $i = 30$\degree, and geometry
equal to the likely values in Figure \ref{histoplots}.   Lower panel -- optical depth perpendicular to the disk as a 
function of radius at the same wavelengths, illustrating that at 160\micron\ and longer, the disks
are typically optically thin over most of the radii emitting at those wavelengths.
}
\end{figure}

It is well understood that the mass of circumstellar disks is best measured at submm or mm wavelengths
where the disks are optically thin except in the very central regions and the dust is virtually all emitting on the 
Rayleigh-Jeans part of its SED.  The longest wavelength observed in our study is 160\micron.
Although the optical depth in the plane of all our model disks is well above unity at 160\micron,
Figure \ref{taufig} shows that for most of the objects which are naturally not observed edge-on, the disks
are vertically optically thin in emission over radial distances where more than half the total flux is emitted.
This figure also shows that the situation at 70\micron\ is more problematic with only $\sim$ 1/3 of the total
flux emitted outside the $\tau = 1$ radius.
This implies that the 160\micron\ flux may be an approximate measure of disk mass, limited
mainly by its relative insensitivity to very cool and very large dust grains and by the fact that 160\micron\ is
not completely into the Rayleigh-Jeans part of the dust emission spectrum.  Figure \ref{mass70160} 
illustrates this possibility by comparing the most probable model disk masses to the observed
70 and 160\micron\ luminosities.  With the exception of the two outliers, both the TWA objects, there does
appear to be a reasonably good correlation between 160\micron\ flux and likely disk mass, as well as even with 70\micron\ luminosity.
No such correlation exists between the total luminosity, for example, and 160\micron\ flux.
Even though our 160\micron\ data already provide good disk mass estimates, further observations of all these objects with ALMA will be extremely important to
determine both the disk fluxes at (sub)mm wavelengths and disk sizes.

\subsection{Host Cloud Correlations?}

The 46 low-mass objects with disks in our study are located in various star-forming
clouds as listed in Table \ref{cloudtable}.  These host clouds have a range of estimated
ages as well as other less quantifiable differences.  For example, the TW Hydra Association is a very low
stellar-density group, while some of the younger regions have a much higher density of young
stars and brown dwarfs in their core.  In any case, a number of figures in our study that distinguish between
these host clouds display no clear correlation of any observed or modeled parameter with
the location of the objects.  We have also investigated a simple grouping of objects into two
bins, those in clouds with estimated ages of order 10 Myr, TWA and Upper Sco,  and the remainder that are significantly younger.
Even with this division which encompasses 13 objects in the ``older'' group, we find no significant
differences in disk properties.  The median probable disk mass in the ``older'' group is 
$log_{10} M_{disk} = -5.0 \pm 0.85$ \msun\ while the median for the younger group is $log_{10} M_{disk} = -4.5 \pm 1.05$ \msun.



\subsection{Comparison With Young Stellar Objects}

In Paper I we made an initial comparison of the disk properties of our first three program objects
with young stars that will eventually burn hydrogen, the T Tauri stars.  Because such stars are
brighter than brown dwarfs, there exists a large quantity of observations and modeling in the
literature to characterize their disks via dust emission, scattering, and even gas emission.  The recent review by \citet{araa11} summarizes the state
of understanding of disks around T Tauri stars. The disk geometries for such stars are
quite similar to the geometries found to be likely fits to the disks around our sample of
very low mass objects within the mutual uncertainties.  The typical surface density gradient we found
of $\Sigma \propto r^{-0.5}$ to $r^{-1.0}$ and typical scale height and flaring mentioned above are certainly
well within the range for models of solar-mass young objects.  It is interesting, though, that if the disk
vertical structure is determined solely by hydrostatic equilibrium with well-mixed gas and dust, lower
central object masses would lead to more flared disks around brown dwarfs than around T Tauri stars \citep{walker04}
which is not
obvious in our results.   This has already been noted by \citet{szucs10} in a comparison of median SED's between
BD's and T Tauri stars.
The most significant clear difference between our sample of sub-stellar objects and T Tauri stars
is the total disk mass inferred from modeling and submm/mm observations.  Disk masses
around T Tauri stars are commonly found to be in the range of $10^{-3} - 10^{-1}$ \msun, well above
our median disk mass of $3 \times 10^{-5}$ \msun.  The most massive disks in our sample are comparable
to the masses of a few previously modeled brown dwarfs (e.g. \citep{bouy08}) and to the low end of
the distribution of disks around T Tauri stars.  The masses of our sample objects are uncertain, but
assuming a median mass $\sim few \times 10^{-2}$ \msun\ for the brown dwarfs, the median ratio of disk mass to central object mass
may be a factor of 3--10 lower than for T Tauri stars.
When the results from the shallow large scale {\it Herschel} surveys are available, it will be interesting to
re-examine this conclusion with the inclusion of brighter brown dwarfs.

\section{Summary}\label{summ}

We have used {\it Herschel/PACS} to observe nearly 50 young sub-stellar and nearly sub-stellar objects with likely circum``stellar'' disks implied by
previous {\it Spitzer} data, and we have increased the number of such objects that are detected
beyond 24\micron\ by a factor of $\sim$5.  We find a wide range in ratios of far-ir to mid-ir flux.
Preliminary modeling of the disks to fit the observed SED's shows that the dust disk geometries are similar
to those of their higher mass counterparts, but the range of masses extends well below the
masses of disks around T Tauri stars, i.e. down below $10^{-6}$ \msun.  This implies that for most of the objects, unless
giant planets have already formed, there is not nearly enough mass to form them.
Our radiative transfer modeling shows that {\it Herschel's} 160\micron\ band is already long enough
to provide reasonable disk mass estimates.
Although our sample selection precludes drawing conclusions about disk frequencies, we have found
no clear dependence of disk mass or geometry  on the age of the star-forming region
where the object is located.

\section{Acknowledgments}
We thank the referee, K. Luhman, for bringing to our attention several spectral type determinations
that we missed and pointing out two of the extragalactic interlopers in our sample.
Support for this work, as part of the NASA Herschel Science Center data analysis funding program, 
was provided by NASA through a contract issued by the Jet Propulsion Laboratory, California 
Institute of Technology to the University
of Texas.  LAC was supported by NASA through the Sagan Fellowship Program.
FM and CP acknowledge support from  ANR (contracts ANR-07-BLAN-0221 and ANR-2010-JCJC-0504-01 ), the
European Commission's 7$^\mathrm{th}$ Framework Program 
(contract PERG06-GA-2009-256513) and 
Programme National de Physique Stellaire (PNPS) of CNRS/INSU, France.
SW acknowledges support by the German Research Foundation
(contract FOR 759). YL acknowledges support by the
German Academic Exchange Service.

This publication makes use of data products from the Wide-field Infrared Survey Explorer, which is a joint project of the University of California, Los Angeles, and the Jet Propulsion Laboratory/California Institute of Technology, funded by the National Aeronautics and Space Administration.

The DENIS project has been partly funded by the SCIENCE and the HCM plans of
the European Commission under grants CT920791 and CT940627.
It is supported by INSU, MEN and CNRS in France, by the State of Baden-W\"urttemberg 
in Germany, by DGICYT in Spain, by CNR in Italy, by FFwFBWF in Austria, by FAPESP in Brazil,
by OTKA grants F-4239 and F-013990 in Hungary, and by the ESO C\&EE grant A-04-046.
Jean Claude Renault from IAP was the Project manager.  Observations were  
carried out thanks to the contribution of numerous students and young 
scientists from all involved institutes, under the supervision of  P. Fouqu\'e,  
survey astronomer resident in Chile.  

\clearpage

\begin{table}[h]
\caption{Location of Observed Young Brown Dwarfs \label{cloudtable}}
\vspace {3mm}
\begin{tabular}{lcccc}
\tableline
\tableline
Cloud &  Number of BD's  & Assumed Distance & Assumed Age  & References \cr
\tableline


Ophiuchus & 3 &  125 pc & 1--3 Myr &  1\\
Taurus & 4 &  140 pc & 1--3 Myr  & 2 \\
Chamaeleon I & 12 & 160 pc & 1--3 Myr & 3\\
Chamaeleon II & 3 & 178 pc & 1--3 Myr & 1\\
Lup I & 1 & 150 pc & 1--3 Myr & 1\\
Lup III & 5 & 200 pc & $\sim$5 Myr & 4 \\
Sigma Ori & 7 & 360 pc & $\sim$3 Myr  & 5 \\
Upper Sco & 11 & 145 pc & $\sim$11 Myr  & 6 \\
TWA & 2 &  54 pc & 8--10 Myr  & 7 \\

\tableline

\end{tabular}
\tablerefs{(1) As adopted by {\it Spitzer c2d}, Evans et al. 2007; (2) Kenyon, G\'omez \& Whitney 2008; (3) Luhman 2008; (4) Comer\'on, F. 2008, 2009; (5) Caballero et al. 2007; (6) Pecaut, Mamajek \& Bubar 2012; (7) Webb et al. 1999 }
\end{table}

\begin{table}[h]
\caption{Observational Parameters \label{obstable}}
\vspace {3mm}
\begin{tabular}{lccc}
\tableline
\tableline
Parameter  & Value  &  Comments \cr
\tableline


OBSID Type &  PACS Mini-Scan-Map &  Two Crossed OBSID's \\
Wavelengths &  70\micron,  160\micron\ & \\
Number of Scan Legs &  8 & \\
Scan Length &  3\arcmin\ & \\
Cross Scan Step & 4\arcsec\ & \\
Scan Angles &  70\degree, 110\degree & Relative to Detector \\
Repetitions &   7  & Per OBSID \\
Peak Intg Time Per Pixel & 504 sec & Per OBSID \\

\tableline

\tableline
\end{tabular}
\end{table}

%








\clearpage



\begin{deluxetable}{llcccccc}
\tabletypesize{\small}
\rotate
\tablecolumns{8}
\tablecaption{Source Summary (Program ID = GT1\_pharve01\_2) \label{aortable}}
\tablewidth{0pt}
\tablehead{
\colhead{Object} &
\colhead{RA/Dec Center (J2000)} &
\colhead{OBSID's} &
\colhead{Obs. Date} &
\colhead{F$_\nu$ 70\micron} &
\colhead{F$_\nu$ 160\micron} &
\colhead{Sp. Type} &
\colhead{Refs\tablenotemark{a} }\\
\colhead{} &
\colhead{h\quad m\quad s\quad \quad \degree\quad \am\quad \as\quad} &
\colhead{} &
\colhead{} &
\colhead{mJy} &
\colhead{mJy} &
}

\startdata
   CFHT-Tau9 & 04 24 26.5 $+$26 49 51 &   1342227059/60 & 2011-08-22 &  9.5 $\pm$  1.2 & $<$ 6 & M6.2 &           14\cr
   KPNO-Tau6 & 04 30 07.2 $+$26 08 21 &    1342227011/2 & 2011-08-21 &  2.3 $\pm$  1.0 & $<$ 5 & M9.0 &        14,15\cr
   KPNO-Tau7 & 04 30 57.2 $+$25 56 40 & 1342227999/8000 & 2011-09-04 &  3.5 $\pm$  1.0 & $<$ 7 & M8.2 &           14\cr
  CFHT-Tau12 & 04 33 09.5 $+$22 46 49 &    1342227013/4 & 2011-08-21 &  1.5 $\pm$  0.7 & $<$ 8 & M6.5 &        14,15\cr
  SOri053825 & 05 38 25.4 $-$02 42 41 &    1342226721/2 & 2011-08-17 &  2.0 $\pm$  1.1 & $<$ 7 & M7.0 &        7,5,6\cr
  SOri053834 & 05 38 33.9 $-$02 45 08 &    1342226723/4 & 2011-08-18 &  7.4 $\pm$  0.9 & 10.1 $\pm$  5.0 &  M4.0  &    33,5,6\cr
  SOri053848 & 05 38 48.2 $-$02 44 01 &    1342226725/6 & 2011-08-18 &  2.7 $\pm$  1.0 & $<$ 4 &      &          5,6\cr
  SOri053855 & 05 38 55.4 $-$02 41 21 &    1342226727/8 & 2011-08-18 &  1.9 $\pm$  1.0 & $<$ 5 & M5.0 &          8,5\cr
  SOri053902 & 05 39 02.0 $-$02 35 01 &    1342228431/2 & 2011-09-07 & 27.9 $\pm$  4.5 & 20.0 $\pm$  4.0 & M2--4  &    8,5,6\cr
      SOri36 & 05 39 26.9 $-$02 36 56 &   1342228429/30 & 2011-09-07 &  1.5 $\pm$  0.8 & $<$ 5 &      &          5,6\cr
  SOri054004 & 05 40 04.5 $-$02 36 42 &    1342228363/4 & 2011-09-09 & $<$1.2 & $<$ 3 &      &          5,6\cr
 SSSPM110209 & 11 02 09.8 $-$34 30 36 &   1342221849/50 & 2011-05-29 &  9.4 $\pm$  0.6 &  8.6 $\pm$  1.8 & M8.5 &     30,13,32\cr
   2MJ110703 & 11 07 03.7 $-$77 24 31 &    1342223480/1 & 2011-06-23 & 24.6 $\pm$  1.2 & 50.1 $\pm$ 12.0 & M7.5 &  18,13,18,19\cr
      ChaHa9 & 11 07 19.2 $-$77 32 52 &    1342223482/3 & 2011-06-23 & 12.2 $\pm$  0.8 & $<$10 & M5.5 &  22,13,21,23\cr
   Cha110804 & 11 08 04.0 $-$77 30 38 &    1342223484/5 & 2011-06-23 & 94.4 $\pm$  1.6 & 84.1 $\pm$  8.0 &      &        13,18\cr
      ISO138 & 11 08 19.0 $-$77 30 41 &  1342218699/700 & 2011-04-16 &  3.4 $\pm$  0.8 & $<$15 & M6.5 &  23,13,18,21\cr
      ISO143 & 11 08 22.3 $-$77 30 28 &  1342218699/700 & 2011-04-16 & 13.3 $\pm$  0.9 & 10.2 $\pm$  2.0 & M5.0 &  23,13,21,22\cr
      ChaHa6 & 11 08 40.2 $-$77 34 17 &    1342223486/7 & 2011-06-34 & 14.4 $\pm$  1.0 & $<$10 & M5.8 &     22,13,23\cr
         T37 & 11 08 50.9 $-$76 25 14 &    1342223468/9 & 2011-06-22 &  2.9 $\pm$  0.9 & $<$ 3 & M5.2 &     23,13,22\cr
      ISO165 & 11 08 55.0 $-$76 32 41 &    1342223470/1 & 2011-06-22 & 42.5 $\pm$  0.8 & 30.0 $\pm$ 10.0 & M5.5 &  23,13,18,21\cr
      Cha726 & 11 09 52.2 $-$76 39 13 &    1342223474/5 & 2011-06-23 & 10.4 $\pm$  1.3 & $<$30 & M6.2 &  22,13,21,23\cr
       OTS44 & 11 10 11.4 $-$76 32 13 &    1342223472/3 & 2011-06-22 &  7.0 $\pm$  1.0 & $<$10 & M9.5 &           24\cr
      ISO252 & 11 10 41.4 $-$77 20 48 &    1342223478/9 & 2011-06-23 &  8.7 $\pm$  0.7 & $<$10 & M6.0 &  23,13,18,21\cr
   2MJ111145 & 11 11 45.3 $-$76 36 50 &    1342223476/7 & 2011-06-23 &  3.0 $\pm$  1.0 & $<$10 & M8.0 &  20,13,18,21\cr
   2MJ120733 & 12 07 33.4 $-$39 32 54 &    1342202557/8 & 2010-08-10 &  7.1 $\pm$  0.7 & $<$ 7 & M8.0 &  31,13,27,29\cr
\tablebreak
    Allers-2 & 12 58 06.7 $-$77 09 09 &    1342224206/7 & 2011-07-14 &  3.4 $\pm$  0.7 & $<$ 3 & M6.0 &          2,1\cr
    Allers-6 & 13 07 18.0 $-$77 40 53 &    1342224202/3 & 2011-07-14 &  6.4 $\pm$  0.9 & $<$12 & M5.0 &          2,1\cr
    Allers-7 & 13 08 27.2 $-$77 43 24 &    1342224204/5 & 2011-07-14 &  6.0 $\pm$  0.9 & $<$ 7 & M6.0 &          2,1\cr
   Allers-19 & 15 44 57.9 $-$34 23 39 &    1342226701/2 & 2011-08-17 &  2.2 $\pm$  0.8 & $<$ 8 & M4.5 &          2,1\cr
   usd155556 & 15 55 56.0 $-$20 45 18 &    1342227138/9 & 2011-08-23 & 12.2 $\pm$  0.7 & 12.0 $\pm$  2.5 & M6.5 &      26,4,25\cr
   usd155601 & 15 56 01.0 $-$23 38 08 &    1342227132/3 & 2011-08-23 &  6.5 $\pm$  0.8 & 11.4 $\pm$  2.1 & M6.5 &      26,4,25\cr
     usco128 & 15 59 11.2 $-$23 37 59 &    1342227134/5 & 2011-08-23 &  5.3 $\pm$  0.7 &  4.1 $\pm$  1.2 & M7.0 &      28,4,25\cr
     usco112 & 16 00 26.6 $-$20 56 32 &    1342227140/1 & 2011-08-23 &  1.6 $\pm$  0.8 & $<$ 7 & M5.5 &      28,4,25\cr
      usco55 & 16 02 45.6 $-$23 04 50 &    1342227136/7 & 2011-08-23 & 11.1 $\pm$  0.9 &  6.0 $\pm$  1.2 & M5.5 &      28,4,25\cr
   DENIS1603 & 16 03 34.7 $-$18 29 30 &    1342227831/2 & 2011-09-03 &  1.6 $\pm$  0.9 & $<$ 5 & M5.5 &          3,4\cr
   usd160603 & 16 06 03.9 $-$20 56 45 &    1342227142/3 & 2011-08-23 &  3.2 $\pm$  1.0 &  3.8 $\pm$  3.3 & M7.5 &      26,4,25\cr
   2MJ160737 & 16 07 37.7 $-$39 21 39 &    1342227813/4 & 2011-09-02 & 14.4 $\pm$  0.9 & 13.2 $\pm$  8.5 & M5.8 &         9,11\cr
   2MJ160814 & 16 08 15.0 $-$38 57 14 &    1342227825/6 & 2011-09-03 & 66.8 $\pm$  1.5 & 49.0 $\pm$  9.0 & M4.8 &         9,11\cr
 IRACJ160828 & 16 08 28.1 $-$39 13 09 &    1342227817/8 & 2011-09-02 &  5.5 $\pm$  1.0 & $<$ 7 & M5.0 &   12,9,13,16\cr
   2MJ160859 & 16 08 59.5 $-$38 56 28 &    1342227823/4 & 2011-09-03 &  7.4 $\pm$  0.9 & 11.8 $\pm$  5.0 & M8.0 &         9,11\cr
   usd160958 & 16 09 58.5 $-$23 45 19 &    1342227146/7 & 2011-08-23 &  4.1 $\pm$  0.7 & $<$ 4 & M6.5 &      26,4,25\cr
   usd161005 & 16 10 05.4 $-$19 19 36 &   1342227829/30 & 2011-09-03 &  4.3 $\pm$  0.8 &  7.5 $\pm$  1.5 & M7.0 &      26,4,25\cr
   usd161833 & 16 18 33.2 $-$25 17 50 &    1342227827/8 & 2011-09-03 &  9.0 $\pm$  1.5 & 12.0 $\pm$  7.0 & M6.0 &      26,4,25\cr
   usd161939 & 16 19 39.8 $-$21 45 35 &    1342227835/6 & 2011-09-03 & 15.3 $\pm$  0.6 &  4.8 $\pm$  3.0 & M7.0 &      26,4,25\cr
  Allers-8 & 16 21 42.0 $-$23 13 43 &    1342227837/8 & 2011-09-03 & 34.3 $\pm$  0.9 & 41.5 $\pm$ 11.0 & M3.0 &          2,1\cr
  Allers-13 & 16 22 45.0 $-$23 17 13 &   1342227839/40 & 2011-09-03 & 12.0 $\pm$  3.5 & $<$15 & M6.0 &          2,1\cr
   CFHTWIR96 & 16 27 40.8 $-$24 29 01 &    1342227841/2 & 2011-09-03 &  5.0 $\pm$  20 & 130 $\pm$ 200 & M8.2 &           17\cr

\tablebreak
\enddata
\tablenotetext{a}{First reference listed is for spectral type if measured; additional references are for various $\lambda <$ 40\micron\ photometry} 
\tablerefs{(1) Allers et al. 2006; (2) Gully-Santiago, Allers \& Jaffe 2012; (3) Bouy et al. 2007; (4) Wright et al. 2010 (WISE); (5) Caballero et al. 2007; (6) Hernandez et al. 2007; (7) Rigliaco et al. 2011; (8) Caballero et al. 2008; (9) Allen et al. 2007; (10) Comer\'on 2011; (11) Comer\'on, Spezzi \& L\'opez Mart\'i 2009; (12) L\'opez Mart\'i, Eisl\"offel \& Mundt 2005; (13) Skrutskie et al. (2MASS) 2006; (14) Guieu et al. 2007; (15) Rebull et al. 2010; (16) M\'erin et al. 2008; (17) Alves de Oliveira et al. 2010; (18) Luhman et al. 2008; (19) Luhman \& Muench 2008; (20) Luhman 2007; (21) L\'opez Mart\'i et al. 2004; (22) Damjanov et al. 2007; (23) Luhman et al. 2004; (24) Luhman et al. 2005; (25) Scholz et al. 2007; (26) Mart\'in, Delfosse \& Guieu 2004; (27) DENIS; (28) Ardila, Mart\'in \& Basri 2000; (29) Riaz, Gizis \& Hmiel 2006; (30) Scholz et al. 2005; (31) Gizis 2002; (32) Luhman et al. 2010; (33) Cody \& Hillenbrand 2011}
\end{deluxetable}

\clearpage

\clearpage

\end{document}